\input harvmac
\input amssym

\def\unit{\relax{\rm 1\kern-.26em I}}
\def\nada{\relax{\rm 0\kern-.30em l}}
\def\tilde{\widetilde}



\def\det{{\rm det}}
\def\CP{{\cal P}}
\noblackbox
\def\IL{\relax{\rm I\kern-.18em L}}
\def\IH{\relax{\rm I\kern-.18em H}}
\def\IR{\relax{\rm I\kern-.18em R}}
\def\IC{\relax\hbox{$\inbar\kern-.3em{\rm C}$}}
\def\IZ{\relax\ifmmode\mathchoice
{\hbox{\cmss Z\kern-.4em Z}}{\hbox{\cmss Z\kern-.4em Z}}
{\lower.9pt\hbox{\cmsss Z\kern-.4em Z}} {\lower1.2pt\hbox{\cmsss
Z\kern-.4em Z}}\else{\cmss Z\kern-.4em Z}\fi}
\def\CM {{\cal M}}
\def\CN {{\cal N}}
\def\CR {{\cal R}}
\def\CD {{\cal D}}

\def\CP {{\cal P }}
\def\CL {{\cal L}}

\def\CO {{\cal O}}

\def\CE {{\cal E}}

\def\CH {{\cal H}}
\def\CC {{\cal C}}
\def\CB {{\cal B}}
\def\CS {{\cal S}}
\def\CA{{\cal A}}
\def\CK{{\cal K}}
\def\cO{{\cal O}}
\def\CO {{\cal O}}

\def\CP {{\cal P }}
\def\CQ {{\cal Q }}
\def\CE{{\cal E }}

\def\CS {{\cal S }}

\def\det{{\rm det}}
\def\Tr{{\rm Tr\,}}

\font\manual=manfnt \def\dbend{\lower3.5pt\hbox{\manual\char127}}

\def\IZ{\relax\ifmmode\mathchoice
{\hbox{\cmss Z\kern-.4em Z}}{\hbox{\cmss Z\kern-.4em Z}}
{\lower.9pt\hbox{\cmsss Z\kern-.4em Z}} {\lower1.2pt\hbox{\cmsss
Z\kern-.4em Z}}\else{\cmss Z\kern-.4em Z}\fi}
\def\half {{1\over 2}}

\def\bar{\overline}
\def\CS{{\cal S}}
\def\CH{{\cal H}}

\def\rt2{\sqrt{2}}
\def\irt2{{1\over\sqrt{2}}}

\def\hat{\widehat}
\def\slashchar#1{\setbox0=\hbox{$#1$}           
   \dimen0=\wd0                                 
   \setbox1=\hbox{/} \dimen1=\wd1               
   \ifdim\dimen0>\dimen1                        
      \rlap{\hbox to \dimen0{\hfil/\hfil}}      
      #1                                        
   \else                                        
      \rlap{\hbox to \dimen1{\hfil$#1$\hfil}}   
      /                                         
   \fi}

\def\foursqr#1#2{{\vcenter{\vbox{
    \hrule height.#2pt
    \hbox{\vrule width.#2pt height#1pt \kern#1pt
    \vrule width.#2pt}
    \hrule height.#2pt
    \hrule height.#2pt
    \hbox{\vrule width.#2pt height#1pt \kern#1pt
    \vrule width.#2pt}
    \hrule height.#2pt
        \hrule height.#2pt
    \hbox{\vrule width.#2pt height#1pt \kern#1pt
    \vrule width.#2pt}
    \hrule height.#2pt
        \hrule height.#2pt
    \hbox{\vrule width.#2pt height#1pt \kern#1pt
    \vrule width.#2pt}
    \hrule height.#2pt}}}}
\def\psqr#1#2{{\vcenter{\vbox{\hrule height.#2pt
    \hbox{\vrule width.#2pt height#1pt \kern#1pt
    \vrule width.#2pt}
    \hrule height.#2pt \hrule height.#2pt
    \hbox{\vrule width.#2pt height#1pt \kern#1pt
    \vrule width.#2pt}
    \hrule height.#2pt}}}}
\def\sqr#1#2{{\vcenter{\vbox{\hrule height.#2pt
    \hbox{\vrule width.#2pt height#1pt \kern#1pt
    \vrule width.#2pt}
    \hrule height.#2pt}}}}

\def\figin{\epsfcheck\figin}\def\figins{\epsfcheck\figins}
\def\epsfcheck{\ifx\epsfbox\UnDeFiNeD
\message{(NO epsf.tex, FIGURES WILL BE IGNORED)}
\gdef\figin##1{\vskip2in}\gdef\figins##1{\hskip.5in}
\else\message{(FIGURES WILL BE INCLUDED)}%
\gdef\figin##1{##1}\gdef\figins##1{##1}\fi}
\def\DefWarn#1{}
\def\figinsert{\goodbreak\midinsert}
\def\ifig#1#2#3{\DefWarn#1\xdef#1{fig.~\the\figno}
\writedef{#1\leftbracket fig.\noexpand~\the\figno}%
\figinsert\figin{\centerline{#3}}\medskip\centerline{\vbox{\baselineskip12pt
\advance\hsize by -1truein\noindent\footnotefont{\bf
Fig.~\the\figno:\ } \it#2}}
\bigskip\endinsert\global\advance\figno by1}

%
%







\input TXSruled.tex


\thicksize=1pt
\def\tstrut{\vrule height 2.5ex depth .8ex width 0pt}
\def\Table#1#2{%
   \setbox0=\hbox to 1.5in{\hfil\bf Table~#1:~}%
   \par\hangindent1.5in\hangafter1%
   \noindent\box0 #2%
}%

\lref\NawataUN{
  S.~Nawata,
  ``Localization of N=4 Superconformal Field Theory on $S^1 \times S^3$ and Index,''
JHEP {\bf 1111}, 144 (2011).
[arXiv:1104.4470 [hep-th]].
}

\lref\WittenGF{
  E.~Witten,
  ``On S duality in Abelian gauge theory,''
Selecta Math.\  {\bf 1}, 383 (1995).
[hep-th/9505186].
}

\lref\MoorePC{
  G.~W.~Moore and E.~Witten,
  ``Integration over the u plane in Donaldson theory,''
Adv.\ Theor.\ Math.\ Phys.\  {\bf 1}, 298 (1997).
[hep-th/9709193].
}

\lref\ArdehaliXLA{
  A.~Arabi Ardehali, J.~T.~Liu and P.~Szepietowski,
  ``The shortened KK spectrum of IIB supergravity on $Y^{p,q}$,''
JHEP {\bf 1402}, 064 (2014).
[arXiv:1311.4550 [hep-th]].
}

\lref\KimWB{
  S.~Kim,
  ``The Complete superconformal index for N=6 Chern-Simons theory,''
Nucl.\ Phys.\ B {\bf 821}, 241 (2009), [Erratum-ibid.\ B {\bf 864}, 884 (2012)].
[arXiv:0903.4172 [hep-th]].
}

\lref\OsbornQU{
  H.~Osborn,
  ``N=1 superconformal symmetry in four-dimensional quantum field theory,''
Annals Phys.\  {\bf 272}, 243 (1999).
[hep-th/9808041].
}

\lref\HofmanAR{
  D.~M.~Hofman, J.~Maldacena,
  ``Conformal collider physics: Energy and charge correlations,''
JHEP {\bf 0805}, 012 (2008).
[arXiv:0803.1467 [hep-th]].
}

\lref\XieJC{
  D.~Xie and P.~Zhao,
  ``Central charges and RG flow of strongly-coupled N=2 theory,''
JHEP {\bf 1303}, 006 (2013).
[arXiv:1301.0210].
}

\lref\GaiottoXA{
  D.~Gaiotto, L.~Rastelli and S.~S.~Razamat,
  ``Bootstrapping the superconformal index with surface defects,''
[arXiv:1207.3577 [hep-th]].
}

\lref\GaddeUV{
  A.~Gadde, L.~Rastelli, S.~S.~Razamat and W.~Yan,
  ``Gauge Theories and Macdonald Polynomials,''
Commun.\ Math.\ Phys.\  {\bf 319}, 147 (2013).
[arXiv:1110.3740 [hep-th]].
}

\lref\ArgyresXN{
  P.~C.~Argyres, M.~R.~Plesser, N.~Seiberg and E.~Witten,
  ``New N=2 superconformal field theories in four-dimensions,''
Nucl.\ Phys.\ B {\bf 461}, 71 (1996).
[hep-th/9511154].
}

\lref\ArgyresJJ{
  P.~C.~Argyres and M.~R.~Douglas,
  ``New phenomena in SU(3) supersymmetric gauge theory,''
Nucl.\ Phys.\ B {\bf 448}, 93 (1995).
[hep-th/9505062].
}

\lref\HananyUC{
  A.~Hanany and C.~Romelsberger,
  ``Counting BPS operators in the chiral ring of N=2 supersymmetric gauge theories or N=2 braine surgery,''
Adv.\ Theor.\ Math.\ Phys.\  {\bf 11}, 1091 (2007).
[hep-th/0611346].
}

\lref\FortinNQ{
  J.~-F.~Fortin, K.~Intriligator and A.~Stergiou,
  ``Current OPEs in Superconformal Theories,''
JHEP {\bf 1109}, 071 (2011).
[arXiv:1107.1721 [hep-th]].
}

\lref\XieHS{
  D.~Xie,
  ``General Argyres-Douglas Theory,''
JHEP {\bf 1301}, 100 (2013).
[arXiv:1204.2270 [hep-th]].
}

\lref\MinahanFG{
  J.~A.~Minahan and D.~Nemeschansky,
  ``An N=2 superconformal fixed point with E(6) global symmetry,''
Nucl.\ Phys.\ B {\bf 482}, 142 (1996).
[hep-th/9608047].
}

\lref\BuicanICA{
  M.~Buican,
  ``Minimal Distances Between SCFTs,''
JHEP {\bf 1401}, 155 (2014).
[arXiv:1311.1276 [hep-th]].
}

\lref\AharonySX{
  O.~Aharony, J.~Marsano, S.~Minwalla, K.~Papadodimas and M.~Van Raamsdonk,
  ``The Hagedorn - deconfinement phase transition in weakly coupled large N gauge theories,''
Adv.\ Theor.\ Math.\ Phys.\  {\bf 8}, 603 (2004).
[hep-th/0310285].
}

\lref\DolanZH{
  F.~A.~Dolan and H.~Osborn,
  ``On short and semi-short representations for four-dimensional superconformal symmetry,''
Annals Phys.\  {\bf 307}, 41 (2003).
[hep-th/0209056].
}

\lref\RomelsbergerEG{
  C.~Romelsberger,
  ``Counting chiral primaries in N = 1, d=4 superconformal field theories,''
Nucl.\ Phys.\ B {\bf 747}, 329 (2006).
[hep-th/0510060].
}

\lref\KinneyEJ{
  J.~Kinney, J.~M.~Maldacena, S.~Minwalla and S.~Raju,
  ``An Index for 4 dimensional super conformal theories,''
Commun.\ Math.\ Phys.\  {\bf 275}, 209 (2007).
[hep-th/0510251].
}

\lref\FestucciaWS{
  G.~Festuccia and N.~Seiberg,
  ``Rigid Supersymmetric Theories in Curved Superspace,''
JHEP {\bf 1106}, 114 (2011).
[arXiv:1105.0689 [hep-th]].
}

\lref\ArdehaliXYA{
  A.~A.~Ardehali, J.~T.~Liu and P.~Szepietowski,
  ``$1/N^2$ corrections to the holographic Weyl anomaly,''
JHEP {\bf 1401}, 002 (2014).
[arXiv:1310.2611 [hep-th]].
}

\lref\MinahanCJ{
  J.~A.~Minahan and D.~Nemeschansky,
  ``Superconformal fixed points with E(n) global symmetry,''
Nucl.\ Phys.\ B {\bf 489}, 24 (1997).
[hep-th/9610076].
}

\lref\GaiottoWE{
  D.~Gaiotto,
  ``N=2 dualities,''
JHEP {\bf 1208}, 034 (2012).
[arXiv:0904.2715 [hep-th]].
}

\lref\ArgyresCN{
  P.~C.~Argyres and N.~Seiberg,
  ``S-duality in N=2 supersymmetric gauge theories,''
JHEP {\bf 0712}, 088 (2007).
[arXiv:0711.0054 [hep-th]].
}

\lref\CachazoRY{
  F.~Cachazo, M.~R.~Douglas, N.~Seiberg and E.~Witten,
  ``Chiral rings and anomalies in supersymmetric gauge theory,''
JHEP {\bf 0212}, 071 (2002).
[hep-th/0211170].
}

\lref\ArgyresTQ{
  P.~C.~Argyres and J.~R.~Wittig,
  ``Infinite coupling duals of N=2 gauge theories and new rank 1 superconformal field theories,''
JHEP {\bf 0801}, 074 (2008).
[arXiv:0712.2028 [hep-th]].
}

\lref\Kodaira{
  K.~Kodaira,
  Ann. of Math. 77 (1963) 563; 78 (1963) 1.
}

\lref\ShapereZF{
  A.~D.~Shapere and Y.~Tachikawa,
  ``Central charges of N=2 superconformal field theories in four dimensions,''
JHEP {\bf 0809}, 109 (2008).
[arXiv:0804.1957 [hep-th]].
}

\lref\BenvenutiPQ{
  S.~Benvenuti, A.~Hanany and N.~Mekareeya,
  ``The Hilbert Series of the One Instanton Moduli Space,''
JHEP {\bf 1006}, 100 (2010).
[arXiv:1005.3026 [hep-th]].
}

\lref\ClossetVRA{
  C.~Closset, T.~T.~Dumitrescu, G.~Festuccia and Z.~Komargodski,
  ``The Geometry of Supersymmetric Partition Functions,''
JHEP {\bf 1401}, 124 (2014).
[arXiv:1309.5876 [hep-th]].
}

\lref\RattazziPE{
  R.~Rattazzi, V.~S.~Rychkov, E.~Tonni and A.~Vichi,
  ``Bounding scalar operator dimensions in 4D CFT,''
JHEP {\bf 0812}, 031 (2008).
[arXiv:0807.0004 [hep-th]].
}

\lref\PolandEY{
  D.~Poland, D.~Simmons-Duffin and A.~Vichi,
  ``Carving Out the Space of 4D CFTs,''
JHEP {\bf 1205}, 110 (2012).
[arXiv:1109.5176 [hep-th]].
}

\lref\BeemSZA{
  C.~Beem, M.~Lemos, P.~Liendo, W.~Peelaers, L.~Rastelli and B.~C.~van Rees,
  ``Infinite Chiral Symmetry in Four Dimensions,''
[arXiv:1312.5344 [hep-th]].
}

\lref\AsselPAA{
  B.~Assel, D.~Cassani and D.~Martelli,
  ``Localization on Hopf surfaces,''
[arXiv:1405.5144 [hep-th]].
}

\lref\GukovYA{
  S.~Gukov, C.~Vafa and E.~Witten,
  ``CFT's from Calabi-Yau four folds,''
Nucl.\ Phys.\ B {\bf 584}, 69 (2000), [Erratum-ibid.\ B {\bf 608}, 477 (2001)].
[hep-th/9906070].
}

\lref\CecottiFI{
  S.~Cecotti, A.~Neitzke and C.~Vafa,
  ``R-Twisting and 4d/2d Correspondences,''
[arXiv:1006.3435 [hep-th]].
}

\lref\SpiridonovWW{
  V.~P.~Spiridonov and G.~S.~Vartanov,
  ``Elliptic hypergeometric integrals and 't Hooft anomaly matching conditions,''
JHEP {\bf 1206}, 016 (2012).
[arXiv:1203.5677 [hep-th]].
}

\lref\BeemKKA{
  C.~Beem, L.~Rastelli and B.~C.~van Rees,
  ``W Symmetry in six dimensions,''
[arXiv:1404.1079 [hep-th]].
}

\lref\ShapereXR{
  A.~D.~Shapere and C.~Vafa,
  ``BPS structure of Argyres-Douglas superconformal theories,''
[hep-th/9910182].
}

\lref\Arnold{
 V.~I.~Arnold, S.~M.~Gusein-Zade and A.~N.~Var$\check{\rm e}$ncko,
 ``Singularities of Differentiable Maps'' (Boston: Birkh\"auser, 1988).
}

\lref\DiPietroBCA{
  L.~Di Pietro and Z.~Komargodski,
  ``Cardy Formulae for SUSY Theories in d=4 and d=6,''
[arXiv:1407.6061 [hep-th]].
}

\lref\DobrevQV{
  V.~K.~Dobrev and V.~B.~Petkova,
  ``All Positive Energy Unitary Irreducible Representations of Extended Conformal Supersymmetry,''
Phys.\ Lett.\ B {\bf 162}, 127 (1985).
}

\lref\DobrevVH{
  V.~K.~Dobrev and V.~B.~Petkova,
  ``On The Group Theoretical Approach To Extended Conformal Supersymmetry: Classification Of Multiplets,''
Lett.\ Math.\ Phys.\  {\bf 9}, 287 (1985).
}

\lref\DobrevQZ{
  V.~K.~Dobrev and V.~B.~Petkova,
  ``Group Theoretical Approach to Extended Conformal Supersymmetry: Function Space Realizations and Invariant Differential Operators,''
Fortsch.\ Phys.\  {\bf 35}, 537 (1987).
}

\Title{\vbox{\baselineskip12pt\hbox{RU-NHETC-2014-12}\hbox{QMUL-PH-14-15}
		}}{Constraints on Chiral Operators in $\CN=2$ SCFTs}

\centerline{Matthew Buican,$^1$\footnote{$^{\dagger}$}
{buican@physics.rutgers.edu} Takahiro Nishinaka,$^1$\footnote{$^{*}$}{nishinaka@physics.rutgers.edu} and Constantinos Papageorgakis$^{1,2}$\footnote{$^{\diamond}$}{c.papageorgakis@qmul.ac.uk}}
\smallskip
\bigskip
\centerline{$^1${\it NHETC and Department of Physics and
Astronomy}}\vskip -.04in
\centerline{{\it Rutgers University, Piscataway, NJ 08854, USA}}
\smallskip
\centerline{$^2${\it CRST and School of Physics and Astronomy}}\vskip -.04in
\centerline{{\it Queen Mary University of London, E1 4NS, UK}}

\vskip .85cm \noindent We study certain higher-spin chiral operators in $\CN=2$ superconformal field theories (SCFTs). In
Lagrangian theories, or in theories related to Lagrangian theories by
generalized Argyres--Seiberg--Gaiotto duality (\lq\lq type A" theories
in our classification), we give a simple superconformal representation
theory proof that such operators do not exist. This argument is
independent of the details of the superconformal index.
We then use the index to show that if a theory
is not of type A but has an $\CN=2$-preserving deformation by a relevant operator
that takes it to a theory of this type in the infrared, the
ultraviolet theory cannot have these higher-spin operators either. 
As an application of this discussion,
we give a simple prescription to extract the $2a-c$ conformal anomaly
directly from the superconformal index. We also comment on how this
procedure works in the holographic limit.

\Date{November 2014}

\newsec{Introduction}
In the space of quantum field theories (QFTs), conformal field theories (CFTs) form a special subspace of enhanced symmetry. The resulting conformal symmetry gives rise to important simplifications. For instance, conformal invariance allows us to describe a CFT by a tightly constrained set of data: its spectrum and operator product expansion (OPE) coefficients.

While it is usually difficult to solve a given CFT (by which we mean
to derive its spectrum and OPE coefficients), one can use general
principles to restrict the space of allowed CFTs. For example, one can
study the constraints imposed by associativity of the OPE and find
bounds on operator dimensions and OPE coefficients; see, e.g.,
\refs{\RattazziPE, \PolandEY}. The restrictions on the space of
superconformal field theories (SCFTs) are potentially even more powerful; see, e.g., \refs{\BeemSZA,\BeemKKA} and references therein.

In this note, we describe new constraints (not derived from associativity of the OPE) on the operator spectra of three very broad (and overlapping) classes of four-dimensional $\CN=2$ SCFTs:

\smallskip
\noindent
\item{\bf (A)} Theories, $\CT$, that have Lagrangian descriptions or theories related to Lagrangian SCFTs by generalized Argyres--Seiberg--Gaiotto duality \refs{\ArgyresCN,\GaiottoWE}: in the latter case, we gauge a global symmetry $G$ of $\CT$ in an exactly marginal fashion (adding additional matter or additional interacting SCFTs charged under the gauge group as necessary) and find a dual description in terms of a Lagrangian.

\smallskip
\noindent
\item{\bf (B)} $\CN=2$ SCFTs with a Coulomb branch.

\smallskip
\noindent
\item{\bf (C)} $\CN=2$ SCFTs admitting deformations by relevant couplings that trigger $\CN=2$-preserving renormalization group (RG) flows to Lagrangian theories (or, more generally, theories of type A) in the infrared (IR).

\smallskip
\noindent
Although it is clear that many theories of type A are also theories of class $\CS$ (the $T_N$ theories are a prototypical set of examples), we will not assume that this is the case in general. Furthermore, as far as we are aware, all known $\CN=2$ SCFTs (whether they are in class  $\CS$ or not) are of type B, with the exception of theories consisting of free hypermultiplets.  
Note that one class of type B theories not of type A is the set of
$(G,G')$  (generalized) Argyres--Douglas (AD) SCFTs
\refs{\ArgyresXN\ArgyresJJ\CecottiFI-\XieHS}.\foot{This statement is
rigorously true for theories with at least one chiral operator with
non-integer scaling dimension.} Finally, the class of
theories of type C is also very broad. For example, it includes all
the $(G, G')$ AD SCFTs.
In fact, it is not immediately clear to us if there are {\it any} theories that are of type B but not of type A or of type C.

The main claims of our paper are:

\smallskip
\noindent
\item{1.} Theories of type A and C do not have a certain class of higher-spin chiral (and anti-chiral) primaries.\foot{An argument against the existence of such operators in certain theories of class $\CS$ was given in \GaddeUV.}

\smallskip
\noindent
\item{2.} If these operators are present in theories of type B, then they necessarily satisfy a non-trivial set of operator relations we will describe in detail.

\smallskip
\noindent
\item{3.} The pole structure of a particular chiral limit of the superconformal index encodes the $2a-c$ conformal anomaly in a very broad class of $\CN=2$ theories and, possibly, all $\CN=2$ SCFTs.

\smallskip
\noindent
Our operators of interest in items 1, 2 above are defined by the shortening conditions
\eqn\Coulombsector{
\left[\tilde \CQ_{I\dot\alpha},
\CO_{\alpha_1\cdot\cdot\cdot\alpha_{2j_1}}\right\}=0~, \  \ \ I=1,2~,
}
where we have a commutator or anti-commutator depending on whether the $SU(2)_1$ Lorentz spin, $j_1$, is even or odd, and $I$ is an $SU(2)_R$ index labeling the two sets of Poincar\'e supercharges. An interesting subset of the operators we study satisfy an additional constraint
\eqn\Dzerocons{\eqalign{
&\left[ \CQ^{I\alpha},
\CO_{\alpha\alpha_1\cdot\cdot\cdot\alpha_{2j_1-1}}\right\}=0~, \ \ \
j_1>0~,\ \ \ I=1,2~,\cr
&\left\{\CQ^{I\alpha},\left[\CQ^J_{\alpha},\CO\right]\right\}=0~, \ \
\ j_1=0~,\ \ \ I, J=1,2~.
}}
When $j_1=0$, the constraint in \Dzerocons\ implies that $\CO$ is a
free scalar satisfying the Klein-Gordon equation (it is the $\CN=2$
primary for a multiplet that contains a free $U(1)$ field
strength). In non-abelian Lagrangian theories, with vector multiplets
involving scalars $\Phi$ in the adjoint of the gauge group, the
$j_1=0$ primaries that satisfy \Coulombsector\ but not \Dzerocons\
have the form $\CO=\Tr\Phi^k$ (with $k\ge2$). Hereafter we will refer to $\CN=2$
primaries with $j_1>0$ satisfying \Coulombsector\ as \lq\lq exotic
chiral primaries."\foot{In the classification of \DolanZH, operators
satisfying \Coulombsector\ and not \Dzerocons\ are denoted as $\bar
\CE_{r(j_1,0)}$. Operators satisfying both \Coulombsector\ and
\Dzerocons\ are denoted by $\bar D_{0(j_1,0)}$.}

In some sense it is surprising that exotic chiral primaries do not
exist in such broad classes of theories.\foot{For example, in $\CN=1$
theories it is common to find chiral primaries that have non-zero spin
\CachazoRY.} For example, operators satisfying \Coulombsector\ but not \Dzerocons\ with $j_1=0$ are ubiquitous in all known $\CN=2$ SCFTs (with the exception of free hypermultiplet theories): their vevs parameterize the Coulomb branch, and so it might have been natural to imagine that exotic chiral operators describe some other geometrical aspects of theories of type A and C. In order to determine if such operators exist at all, it would be interesting to understand whether there are theories of type B that are not of type A or type C (or, if there are any theories outside of our classification), and, if so, whether one can use the operator relations we derive to prove that these operators do not exist in such theories. One might also attempt to study such operators using the conformal bootstrap.

Let us briefly describe the plan of the rest of the paper. In the next section, we discuss certain short multiplets of the four-dimensional $\CN=2$ superconformal algebra, highlighting our multiplets of interest. We then give a general argument that theories of type A do not have exotic chiral operators. Our proof relies only on superconformal representation theory (and the dynamical assumption of a duality with a Lagrangian theory). In order to set the stage for our discussion of theories of type B and C, we then introduce the superconformal index and a particularly useful limit of the index---the \lq\lq Coulomb branch" or \lq\lq chiral $U(1)_R$" index---that captures only contributions from operators of the general type defined in \Coulombsector\ and \Dzerocons. We then study a special limit of the chiral $U(1)_R$ index and use the structure of the resulting divergences to argue that if theories of type B have exotic chiral operators, then they necessarily satisfy certain operator relations. In the next section we move on to theories of type C and show that exotic chiral operators do not exist in such theories. Finally, we conclude with an application of our results to an index-based computation of the $2a-c$ conformal anomaly.

\newsec{$\CN=2$ SCFT Generalities and the Chiral $U(1)_R$ Sector} Any
four-dimensional $\CN=2$ SCFT, $\CT$, has a conserved and traceless
stress tensor along with two conserved supercurrents and conserved
currents for the $SU(2)_R\times U(1)_R$ superconformal $R$
symmetry. Such theories typically admit a zoo of short representations
(note that we will refer to \lq\lq semi-short" representations as
short representations) with $\CN=2$ primaries that are annihilated by
certain combinations of the Poincar\'e supercharges \refs{\DolanZH,\DobrevQV\DobrevVH-\DobrevQZ}. One
universal short representation that is present in any $\CT$ is the
multiplet $J$ that contains the aforementioned stress tensor,
supercurrents, and $U(1)_R\times SU(2)_R$ currents. In the notation of
\refs{\DolanZH,\GaddeUV,\BeemSZA},\foot{We provide a short summary of these
conventions in Appendix A.} the $J$ multiplet is of type $\hat
C_{0(0,0)}$ (it has zero $U(1)_R\times SU(2)_R$ charge, is a Lorentz
scalar, and has scaling dimension two). Superconformal representation
theory allows various short higher spin cousins of this multiplet,
$\hat C_{0(j_1,j_2)}$, that contain conserved higher spin currents and
are therefore not present unless $\CT$ has a sector of free
fields.\foot{All Lagrangian $\CN=2$ SCFTs have a special
limit of measure zero on their conformal manifolds where free fields
emerge, but away from this limit the various $\hat C_{0(j_1,j_2)}$
multiplets combine according to the rules given in \DolanZH\ to become
long multiplets with non-zero anomalous dimension.}

If $\CT$ has a continuous flavor symmetry, $G$, (i.e., a continuous symmetry that commutes with the $\CN=2$ superconformal algebra), then the corresponding spin one symmetry current is a descendant in a short multiplet of type $\hat B_{1}$.\foot{Note that while Lagrangian $\CN=2$ SCFTs necessarily have flavor symmetries, the same is not true in general for interacting theories. For example, the original AD theory in \ArgyresJJ\ does not have any flavor symmetry.} The primary, $L_G^{IJ}$, is a dimension two scalar of $SU(2)_R$ spin one and $U(1)_R$ charge zero satisfying
\eqn\primaryCurr{
L_G^{IJ}=L_G^{JI}~, \ \ \ \left(L_G^{IJ}\right)^{\dagger}=\epsilon_{IK}\epsilon_{JL}L_G^{KL}~, \ \ \ \left[\CQ_{\alpha}^{1},L_G^{11}\right]=\left[\tilde \CQ_{2\dot\alpha},L_G^{11}\right]=0~,
}
where $I, J=1,2$ are $SU(2)_R$ indices. Defining $L_G^A:=-{i\over2}\sigma^{A J}_{\ I}(L_G)^{I}_{J}$, where the $\sigma^A$ are the $SU(2)_R$ Pauli matrices, we see from \primaryCurr\ that $L^3_G$ is the real moment map corresponding to $G$. This fact will be useful for us in our discussion below.

The main focus of our paper is on the set of short multiplets of the $\CN=2$ superconformal algebra whose primaries are annihilated by both sets of anti-chiral Poincar\'e supercharges. Indeed, such operators are necessarily of the type we discussed in \Coulombsector
\eqn\chiralops{
\left[\tilde \CQ_{I\dot\alpha}, \CO^{I_1\cdot\cdot\cdot I_{2R}}_{\alpha_1\cdot\cdot\cdot\alpha_{2j_1};\dot\alpha_1\cdot\cdot\cdot\dot\alpha_{2j_2}}\right\}=0 \ \ \ \Rightarrow \ \ \ j_2=R=0~, \ \ \ E=-r\ge1+j_1~,
}
where $R$ is the $SU(2)_R$ spin, $E$ is the scaling dimension, $r$ is the $U(1)_R$ charge, and the implication in \chiralops\ follows from unitarity. If the operator does not satisfy any additional shortening conditions, it is referred to as being of type $\bar\CE_{r(j_1,0)}$. Note that the inequality in \chiralops\ is saturated if and only if the operator also satisfies \Dzerocons---such operators are referred to as being of type $\bar D_{0(j_1,0)}$. In a slight abuse of terminology, we will refer to the operators in \chiralops\ as constituting the \lq\lq chiral $U(1)_R$" sector of the theory.\foot{One sometimes refers to primaries that have $j_2=0$ as chiral primaries. Here we only refer to operators annihilated by the full set of anti-chiral supercharges as being in the chiral $U(1)_R$ sector of the theory. Note that the superconformal algebra allows other short multiplets charged under $U(1)_R$ but not under $SU(2)_R$ that have $j_2=0$ but are not annihilated by the anti-chiral supercharges (e.g., the $\bar\CC_{0,r(j_1,0)}$ multiplets).} We will be particularly interested in understanding the \lq\lq exotic" chiral $U(1)_R$ sector of the theory---namely, those operators satisfying \chiralops\ with $j_1>0$.

The $U(1)_R$ chiral sector operators have two special properties that will be important in what follows:

\item{\bf (i)} The $\bar\CE_{r(j_1,0)}$ and $\bar D_{0(j_1,0)}$ multiplets cannot combine with other short representations of the $\CN=2$ superconformal algebra to form long representations.

\item{\bf (ii)} The $\bar\CE_{r(j_1,0)}$ and $\bar D_{0(j_1,0)}$ operators are singlets under any flavor symmetries of the theory.

\smallskip
\noindent
The first property follows from the analysis in \DolanZH. We will
prove the second property by contradiction. To that end, let us assume
that the $U(1)_R$ chiral sector operators are charged under some
flavor symmetry. Then, since the $\bar\CE_{r(j_1,0)}$ and $\bar
D_{0(j_1,0)}$ operators are chiral with respect to an
$\CN=1\subset\CN=2$ sub-algebra, it must be the case that the chiral-anti-chiral OPE
includes terms of the type
\eqn\OPE{\eqalign{
\CO_{(a)\alpha_1\cdot\cdot\cdot\alpha_{2j}}(x)\ \CO^{\dagger}_{(\tilde b)\dot\alpha_1\cdot\cdot\cdot\dot\alpha_{2j}}(0)&\supset I_{\alpha_1\cdot\cdot\cdot\alpha_{2j},\dot\alpha_1\cdot\cdot\cdot\dot\alpha_{2j}}\left({1\over x^{2E}}\delta_{a\tilde b}+{1\over x^{2E-2}}T^{\hat A}_{a\tilde b} \cdot L_{\hat A}^3(0)\right)~,\cr I_{\alpha_1\cdot\cdot\cdot\alpha_{2j},\dot\alpha_1\cdot\cdot\cdot\dot\alpha_{2j}}&={1\over(2j)!x^{2j}}\left(\prod_i x_{\alpha_i\dot\alpha_i}+{\rm perm}\right)~,
}}
where $a$ runs over the full set of $\CO$ with the same $\CN=2$ superconformal quantum numbers, $L^3_{\hat A}$ is the real moment map corresponding to a flavor symmetry generator labeled by $\hat A$ (see the discussion below \primaryCurr), and
\eqn\Tdef{
T^{\hat A}_{a\tilde b}={1\over4\pi^2}\gamma^{\hat A\hat B}(t_{\hat B})_a^{\ c}\delta_{c\tilde b}~, \ \ \ \langle L^A_{\hat A}(x)L^B_{\hat B}(0)\rangle={1\over x^4}\delta^{AB}\gamma_{\hat A\hat B}~.
}
The $t_{\hat B}$ matrix in \Tdef\ is the representation matrix that
$\CO_{(a)\alpha_1\cdot\cdot\cdot\alpha_{2j}}$ transforms under, when
acted on by the symmetry generator labeled by $\hat B$. It appears in the OPE
\eqn\currOPE{
j_{\mu, \hat B}(x)\CO_{(a)\alpha_1\cdot\cdot\cdot\alpha_{2j}}(0)\supset-i(t_{\hat B})_a^{\ b}{x_{\mu}\over2\pi^2x^4}\CO_{(b)\alpha_1\cdot\cdot\cdot\alpha_{2j}}(0)~,
}
where $j_{\mu, \hat B}$ is the symmetry current corresponding to the symmetry generator $\hat B$ (note that \currOPE\ is equivalent to the statement that the charge, $Q_{\hat B}\equiv\int d^3xj_{0,\hat B}$, acts on the exotic chiral operator in the way we claim).\foot{We can further justify this statement as follows. Our discussion can be rephrased in terms of the following superconformally covariant three-point function (we refer the reader to \OsbornQU\ for further details on notation and conventions)
\eqn\threept{\eqalign{
\langle\CO_{(a)}^{i}(z_a)\CO_{(\tilde b)}^{\dagger j}(z_{\tilde b})L^3_{\hat C}(z_{\hat C})\rangle&={{I_{(a)}^{i\bar i}}(x_{a\bar {\hat C}},x_{\bar a \hat C}){I_{(\tilde b)}^{j\bar j}}(x_{\tilde b\bar{\hat C}},x_{\bar {\tilde b}{\hat C}})\over x_{\bar a \hat C}^{2\bar q_{a}}x_{\bar {\hat C} a}^{2q_a}x_{\bar {\tilde b} \hat{C}}^{2\bar q_{\tilde b}}x_{\bar{\hat C} \tilde b}^{2q_{\tilde b}}}\cdot t_{\bar i\bar j}(X_{\hat C}, \Theta_{\hat C}, \bar\Theta_{\hat C})~,\cr&={{I_{(a)}^{i\bar i}}(x_{a\bar {\hat C}}){I_{(\tilde b)}^{j\bar j}}(x_{\bar {\tilde b}{\hat C}})\over x_{\bar {\hat C} a}^{2q_a}x_{\bar {\tilde b} \hat{C}}^{2\bar q_{\tilde b}}}\cdot t_{\bar i\bar j}(X_{\hat C}, \Theta_{\hat C}, \bar\Theta_{\hat C})~,
}}
where $i$ and $j$ represent the spins (i.e., $i\sim\alpha_1\cdots\alpha_{2j}$). Note that the second equality follows from the fact that $\bar q_a=q_{\tilde b}=0$ for our chiral operators ($\bar q_a=0$ is equivalent to $E=-r$ in \chiralops; the equation $q_{\tilde b}=0$ follows from the analogous equation for anti-chiral operators) and the fact that the $I^{i\bar i}_{(a)}$ may be constructed from products of the $I_{\alpha\dot\alpha}$ in (3.24) of \OsbornQU\ which only depend on $x_{a\bar{\hat{C}}}$ (and analogously for $I^{j\bar j}_{(\tilde b)}$). 

In this language, to show that the OPE coefficients of the moment maps, $L^3_{\hat C}$, in \OPE\ and the OPE coefficients in \currOPE\ are related as in \Tdef, it is sufficient to show that $t_{\bar i\bar j}$ does not depend on the Grassmann parameters $\Theta_{\hat C}$ and $\bar\Theta_{\hat C}$ (since the current, $j_{\mu, \hat C}$, is a superconformal descendant of the moment map, $L^3_{\hat C}$; indeed, in $\CN=1$ language, we have that $L^3_{\hat C}$ is the primary of a superfield that has the form $L^3_{\hat C}+\cdots-\theta\sigma^{\mu}\bar\theta j_{\mu, \hat C}+\cdots$). To prove this latter statement, we use the fact that $\bar D_{1a}^{\dot\alpha}$ annihilates the LHS of \threept\ to conclude, via (6.1) and (6.2) of \OsbornQU, that $t_{\bar i\bar j}$ is a function only of $\bar X_{\hat C}$ and $\bar\Theta_{\hat C}$. Finally, using the fact that $D_{\tilde b\alpha}^1$ annihilates the LHS of \threept, and again applying (6.1) and (6.2) of \OsbornQU, we find that $t_{\bar i\bar j}$ is independent of $\bar\Theta_{\hat C}$ as well.} Now, since $L^3_{\hat A}$ is part of an $SU(2)_R$ triplet and $\CO_{(a)\alpha_1\cdot\cdot\cdot\alpha_{2j}}$ is an $SU(2)_R$ singlet, we see that $T^{\hat A}_{a\bar b}=0$. As a result, from \Tdef\ and unitarity, $(t_{\hat B})_a^{\ b}=0$, and so the chiral $U(1)_R$ sector operators cannot be charged under any flavor symmetries.

\newsec{Constraints on Exotic Chiral Operators}
In this section we study constraints on the exotic chiral operators we introduced above. In the first subsection, we apply our previous results to conclude that theories of type A do not have such operators. In the next subsection we introduce the superconformal index and a particularly useful limit of it in order to set the stage for our discussion of theories of type B and C.

\subsec{Theories of Type A}
For Lagrangian SCFTs it is straightforward to prove that there are no exotic chiral operators since such theories have exactly marginal gauge couplings that can be tuned to zero. By property {\bf (i)} of the previous section, we see that superconformal representation theory prevents the $\bar\CE_{r(j_1,0)}$ and $\bar D_{0(j_1,0)}$ operators from pairing up to form long representations, and so it suffices to study the free theory. At zero coupling, we immediately see that the only possible chiral operators have the form $\CO=\Tr\Phi^k$, where $\Phi$ is a chiral adjoint, and so $j_1=0$.

Let us now consider a non-Lagrangian theory, $\CT$, that we can deform to a Lagrangian one. We assume that such a deformation does not induce an RG flow, and so it must proceed by an exactly marginal gauging of some global symmetry group of $\CT$, $G_0\subset G_{\CT}$ (see, for example, the appendix of \BuicanICA). In order to ensure that the gauging is exactly marginal, we may have to add another SCFT, $\CT'$, with $G_0\simeq G_0'\subset G_{\CT'}$  ($\CT'$ may be a collection of decoupled SCFTs; note that $\CT'$ may just be a set of free hypermultiplets). In the spirit of \refs{\ArgyresCN,\GaiottoWE} we may then be able to find a duality
\eqn\duality{
\CL \ \ \ \leftrightarrow \ \ \ \CT-G-\CT'~, \ \ \ G={\rm diag}\left(G_0\times G_0'\right)~,
}
where $\CL$ is a Lagrangian theory, and we have identified the gauge group $G$ with the diagonal subgroup of $G_0\times G_0'$. In general, the remaining flavor symmetry, $G_{\CL}$, is just the maximal commuting subgroup with $G$ in $G_{\CT}\times G_{\CT'}$
\eqn\symms{
G_{\CL}=\left\{T\in G_{\CT}\times G_{\CT'}|\ \left[T, \hat T\right]=0 \ \forall\ \hat T\in G\right\}~.
}
In writing \symms, we have used the fact that flavor symmetries are
non-anomalous in $\CN=2$ theories since the moment maps satisfy (see, e.g. \refs{\FortinNQ,\BuicanICA})
\eqn\realmomope{
L_{\hat A}^A(x)L_{\hat B}^B(0)\supset{\kappa_{\hat A\hat B}^{\ \ \ \ \hat C}\over x^2}\epsilon^{AB}_{\ \ C}L_{\hat C}^C(0)~,
}
where $\kappa$ is the (putative) flavor anomaly and $\epsilon$ the
Levi-Civita tensor. In particular, we see that if $A=B=C=3$, then the RHS of \realmomope\ vanishes and the symmetries must be non-anomalous (by the $\CN=1$ arguments in \refs{\FortinNQ,\BuicanICA}).

Let us now study the chiral $U(1)_R$ sector of the theories appearing in \duality. In particular, let us take the limit of zero gauge coupling for the $G$ gauge group, $\tau_G\to i\infty$. In this limit, it is clear that the only effect of the gauging on $\CT$ and $\CT'$ is to eliminate non gauge-invariant operators. However, by property {\bf (ii)} in the previous section, we see that the $\bar\CE_{r(j_1,0)}$ and $\bar D_{0(j_1,0)}$ operators are not charged under $G$. Therefore, such operators persist on the right hand side of \duality, and we must have
\eqn\chiralsectorcomp{
\left\{\bar\CE^{(\CL)}_{r(0,0)}, \bar D_{0(0,0)}^{(\CL)}\right\}=\left\{\bar\CE^{(G)}_{r(0,0)}\right\}\cup\left\{\bar\CE^{(\CT)}_{r(j_1,0)}, \bar D_{0(j_1,0)}^{(\CT)}\right\}\cup\left\{\bar\CE^{(\CT')}_{r(j_1,0)},\bar D_{0(j_1,0)}^{(\CT')}\right\} \ \  \Rightarrow \ \  j_1=j_1'=0~,
}
where the implication that there are no exotic chiral primaries in $\CT$ or $\CT'$ follows from the fact that there are no exotic chiral primaries in the Lagrangian sectors $\CL$ and $G$.

As a simple example, we can consider the original duality of
\ArgyresCN. In that case, $\CL$ is the $SU(3)$ theory with six flavors, $\CT$
is the $E_6$ theory of \MinahanFG, $G=SU(2)$, and $\CT'$ is a
collection of two hypermultiplets (so $G_{\CT}=E_6$, $G_{\CT'}=Sp(2)$,
and $G_{\CL}=U(1)\times SU(6)$). By our above reasoning, we see that
the chiral $U(1)_R$ sector of the pure $E_6$ theory (without gauging
an $SU(2)\subset E_6$) is fully described by a non-exotic dimension
three operator, $\CO$, dual to the $\Tr\Phi^3$ operator in $\CL$. 

Let us again emphasize that while many constructions of the type \duality\ are in class $\CS$, we do not need to assume that this is the case more generally for our above logic to hold. Finally, let us mention that there are large classes of theories in which \duality\ does not hold because (among other possibilities) there are no appropriate symmetries to gauge (such theories are of type B or C but not A). One example is the original AD theory considered in \ArgyresJJ\ (and many of the generalizations in \XieHS). To constrain the existence of chiral exotics in such theories, we will need some more powerful tools which we introduce in the next subsection.

\subsec{The $\CN = 2$ Superconformal Index}
The superconformal index \refs{\RomelsbergerEG,\KinneyEJ} counts short
representations of the superconformal algebra modulo combinations of short
representations that can pair up to form long representations (the
index is therefore invariant under exactly marginal deformations). The
counting is taken with respect to some supercharge, $\CQ$, and is
weighted by fugacities for the symmetries that commute with $\CQ$. In
what follows, we choose the supercharge $\tilde \CQ_{2\dot-}$ (other
choices lead to equivalent constructions of the index) and define the
index to be\foot{We will use the same letter for both
the Cartan generators and the associated charges, hoping that this will not cause
confusion. We review our conventions for the superconformal algebra in Appendix A.}
\eqn\indexdef{\eqalign{
\CI(p,q,t)&=\Tr_{\CH}(-1)^F e^{-\beta
\tilde\delta_{2\dot -}}p^{j_2+j_1}q^{j_2-j_1}t^{R}\left({pq\over t}\right)^{-r}~,\cr
\tilde\delta_{2\dot -}&:=2 \left\{\tilde \CQ_{2\dot-}, \tilde\CS^{2\dot-}\right\}=E-2j_2-2R+r~,
}}
where $p, q, t$ are complex fugacities, $j_{1,2}$ are Cartans for the
$SU(2)_{1,2}$ isometry, $r$ is the superconformal $U(1)_R$ charge, and
$R$ is the Cartan of $SU(2)_R$. In \indexdef, the trace is understood
as being over the full Hilbert space of the theory. However, the only
contributions come from operators in ${\bf R^4}$ (or, equivalently,
the corresponding states on $S^3$) that are annihilated by $\tilde
\CQ_{2\dot-}$ and $ \tilde \CS^{2\dot-}$; such
operators (or states) necessarily have $E=2j_2+2R-r$.\foot{Note that we can
often add various flavor fugacities to \indexdef, but we will not do so in
what follows since our primary interest is in operators that we have
seen are flavor neutral.}

In a Lagrangian theory, the gauge and matter sectors contribute to the index as follows
\eqn\singleFV{\eqalign{
&\CI_{V}^{\rm s.l.}(p,q,t)=-{p\over1-p}-{q\over1-q}+{{pq\over t}-t\over(1-p)(1-q)}~,\cr&\CI^{\rm s.l.}_{H\over2}(p,q,t)={\sqrt{t}-{pq\over\sqrt{t}}\over(1-p)(1-q)}~,
}}
where the subscripts \lq\lq$V$" and \lq\lq${H\over2}$" refer to contributions from an $\CN=2$ vector multiplet and half hypermultiplet respectively; the superscript \lq\lq s.l." stands for \lq\lq single letter" and indicates that we should appropriately exponentiate the corresponding contribution to the index in order to compute the contributions from all short multiplets built out of products of $V$, $H$, and their derivatives. For simplicity, we will specialize our Lagrangian discussion to a $U(1)^N$ gauge theory with free (uncharged) hypermultiplets (the non-abelian and charged-matter generalizations are straightforward) and we will drop all flavor fugacities. In this case, we find
\eqn\Indexabelian{\eqalign{
\CI&=\prod_{a=1}^N {\rm P.E.}(I_{V_a}^{\rm s.l.})\prod_i {\rm P.E.} \left(I_{H_i\over2}^{\rm s.l.}\right)\cr&:=\prod_{a=1}^N\exp\left(\sum_{n=1}^{\infty}{1\over n}I_{V_a}^{\rm s.l.}(p^n,q^n,t^n)\right)\prod_i\exp\left(\sum_{n=1}^{\infty}{1\over n}I_{H_i\over2}^{\rm s.l.}(p^n,q^n,t^n)\right)~,
}}
where the product is over all the abelian $U(1)$'s and matter fields, and \lq\lq P.E." stands for the \lq\lq plethystic exponential."

Via the state-operator map, one can also understand the index as being
equivalent---modulo regularization scheme-dependent pre-factors---to a
partition function for the theory on $S^1\times S^3$ with twisted
boundary conditions for the various fields according to their charges
\refs{\RomelsbergerEG\KinneyEJ\FestucciaWS-\AharonySX}.\foot{In line
with \FestucciaWS, the version of the theory on $S^1\times S^3$ has
additional mass terms, coupling the scalars to the background
curvature.} By assigning a chemical potential to each fugacity, $x =
e^{-\beta v_x}$, we can write \indexdef\ as
\eqn\indexchpot{\eqalign{
  \CI & = \Tr_{\CH } (-1)^F {\rm exp}\left[-\beta( E - 2j_2 - 2R +r +
  v_p(j_1 + j_2 - r)+ v_q (j_2-j_1-r) + v_t(R + r)\right]
\cr &=:  \Tr_{\CH } (-1)^F {\rm exp}\left[-\beta( E + \alpha)\right]~,
}}
where note that $[E + \alpha,\tilde\CQ_{2\dot -}] = 0$. The twisted boundary conditions can be replaced with periodic ones as $x_4 \sim x_4
+ \beta$, by additionally shifting the derivative along the circle \KimWB
\eqn\shift{
\partial_4\to\partial_4 - \alpha =: D_4\;.
}
The corresponding Euclidean path integral, schematically $\int[\CD \Phi]e^{-S_E[\Phi,D_4\Phi]}$,
then counts the same ${1\over 8}$-BPS states as the index. This will
provide us with an interesting alternative perspective below.  Note
that the index is by construction independent of continuous couplings. As a result,
the evaluation of the twisted partition function for the $\CN=1$
chiral multiplets inside the $\CN =2 $ vector or hypermultiplets can
be carried out in the free limit, where it reduces to one-loop
determinant factors for the scalars and fermions
\refs{\KinneyEJ,\AharonySX,\NawataUN}.\foot{For the $\CN=1$ vector
multiplet contributions one needs to take into account gauge field
zero-modes that have to be dealt with exactly.}

It is interesting to point out that there is yet another, more geometric, interpretation
of the index as corresponding (modulo pre-factors) to the partition function of the theory placed on a Hopf
surface, $M_4$. The complex structure and holomorphic bundle moduli of $M_4$
are directly related to the fugacities of the index \ClossetVRA\ (see also \AsselPAA).

\subsec{The Chiral $\hat U(1)_R$ Limit}

Since we will be interested in analyzing theories of type
B---i.e., theories with a Coulomb branch, $\CM_{\CC}$, in ${\bf
R^4}$---it is useful for us to ask how we can see such a moduli space
from the index \Indexabelian\ in the simple case above (see also the general discussion of moduli spaces and the index in \GaiottoXA). In this
example,  $\CM_{\CC}$ is parameterized by the $U(1)_R$-charged vector
multiplet scalars, $\Phi_a$. Therefore, away from the origin of $\CM_{\CC}$, the $U(1)_R$ symmetry is broken. As a result, if we wish to study the effects of such vacua, we should take a limit of fugacities $pq=t$ in \indexdef\ so that the $U(1)_R$ charge does not
enter the index. We define
\eqn\redcoul{
\CI_{\hat U(1)_R}:=\Tr_{\CH}(-1)^F e^{-\beta \tilde\delta_{2\dot -}}p^{j_2+j_1 +
R}q^{j_2-j_1 + R}~, \ \ \ {pq\over t}=1~.
}
Using this index to evaluate the single letter contributions for
vector multiplets and half-hypermultiplets, we find that \singleFV\ becomes
\eqn\singleFVi{\eqalign{
&\CI_{\hat U(1)_R \ V}^{\rm s.l.}(p,q,t)=1~,\cr&\CI^{\rm s.l.}_{\hat U(1)_R\ {H\over2}}(p,q,t)=0~.
}}
The first contribution can be understood as coming from a $\Phi_a$
scalar with no derivatives acting on it. Note that the hypermultiplet
contributions vanish, and our result depends only on
operators in the chiral $U(1)_R$ sector; hence the subscript
\lq\lq$\hat U(1)_R$" in \singleFVi.  Plugging these results into
\Indexabelian, we find unsupressed contributions from all
powers of $\Phi_a$
\eqn\Indexabelianlim{
\CI_{\hat U(1)_R}=\prod_a \exp\left(\sum_{n=1}^{\infty}{1\over n}\right)=\infty~.
}
This divergence describes the opening up of a Coulomb branch.

In fact, this discussion is more general. Indeed, we find the following single letter contributions to the index for any $\CN=2$ SCFT in the above limit:
\eqn\indexcomp{\eqalign{
&\CI^{\rm s.l.}_{\hat U(1)_R\ \CC_{R,r(j_1,j_2)}}=0~,\cr&\CI^{\rm s.l.}_{\hat U(1)_R\ \hat\CC_{R(j_1,j_2)}}=0~,\cr&\CI^{\rm s.l.}_{\hat U(1)_R \ \bar\CE_{r(j_1,0)}}=(-1)^{2j_1}\chi_{j_1}\left(\sqrt{p\over q}\right)~,\cr&\CI^{\rm s.l.}_{\hat U(1)_R\ \bar\CD_{0(j_1,0)}}=(-1)^{2j_1}{(1+pq)\chi_{j_1}\left(\sqrt{p\over q}\right)-\sqrt{pq}\left(\chi_{j_1+{1\over2}}\left(\sqrt{p\over q}\right)+\chi_{j_1-{1\over2}}\left(\sqrt{p\over q}\right)\right)\over(1-p)(1-q)}~,\cr&\CI^{\rm s.l.}_{\hat U(1)_R\ \CD_{0(0,j_2)}}=0~,
}}
where $\chi_{j_1}(x)={x^{2j_1+1}-x^{-(2j_1+1)}\over x-x^{-1}}$ is a
character for the spin $j_1$ representation of $SU(2)_1$. The
subscripts denote contributions from the various short representations
of the $\CN=2$ superconformal algebra. We review the
various properties of these different multiplets in Appendix A. However, suffice it
to say that the only contributions to the index in \indexcomp\ come
from the chiral $U(1)_R$-sector operators $\bar\CE_{r(j_1,0)}$ and
$\bar D_{0(j_1,0)}$ we introduced above.

The divergence of the index in the above limit, as well as the
absence of hypermultiplet contributions, have simple physical
interpretations in terms of the path integral description.  Note that this construction is useful for theories with Lagrangian limits.
In order to proceed, let us first remind the reader that in the
$M_4\simeq S^1\times S^3$ path integral language, for generic
values of chemical potentials, the scalars $\Phi_a$ typically have a
mass (since they couple to curvature). Due to this mass term, the
theory on curved space has a unique vacuum, as opposed to the moduli
space of vacua enjoyed by the theory in flat space. However, for the
limit described in \redcoul, we will find bosonic zero modes in the
$\Phi_a$.

The Lagrangian at quadratic order in the fields includes the
following terms for the vector multiplet scalars
 \eqn\scalars{
  \CL_{S^1\times S^3} \supset \Tr
\bar \Phi \Big( - \Delta_{S^3} - D_{4}^2 + 1\Big) \Phi \;, 
}
where $\Delta_{S^3}$ is the Laplacian on the three-sphere and the
factor of 1 comes from the conformal coupling to the curvature of the
$S^3$, in units of its radius \FestucciaWS. One can read off the scalar mass from \scalars:
\eqn\scalarmass{
  M^2_\phi =(1+\alpha)(1-\alpha) \;.
}
Further focusing on the constant mode ($s$-wave) for the vector multiplet scalars
on $S^1\times S^3$, such that
$j_1= j_2 =0$, $r = -1$ and $R = 0$, produces
\eqn\redmass{
  M^2_{\phi,s} =( 2- v_p -v_q +v_t)(v_p+v_q -v_t) \;.
}
Clearly, in the limit where $pq= t$, or equivalently $v_p +v_q -v_t =
0$ with $\beta$ fixed, this mass goes to zero and the Coulomb branch
opens up. Hence, the divergence in the $\CI_{\hat{U}(1)_R}$ index for
$pq = t $, \Indexabelianlim, can be understood as the result of
integrating over this zero mode in field space. This insight can be
applied more generally. Appropriately tuning the various fugacities
can lead to opening up different flat directions; see \GaiottoXA\ for
a baryonic Higgs-branch example.

We can similarly interpret the absence of hypermultiplet
contributions. In the $ pq =t$ limit, the bosonic and fermionic
contributions to the single-letter hypermultiplet index \singleFVi\
canceled out. In the path integral description, this is suggestive of an
additional supersymmetric cancelation between chiral scalars and
fermions. It can indeed be seen that
\eqn\shiftedcomm{
\lim_{pq \to t}[E + \alpha,\tilde \CQ_{1\dot +}] =
0\;,
}
that is, in the limit of interest the partition function counts ${1
\over 4}$-BPS states annihilated by both $\tilde \CQ_{2\dot -}$ and
$\tilde \CQ_{1\dot+}$, thus excluding the hypermultiplets.

\subsec{The Chiral $U(1)_R$ Limit}

We have seen that the limit of the index \redcoul\ captures
contributions only from Coulomb-branch operators. However, before
proceeding to analyze the chiral $U(1)_R$ sector operators of
interest, we would like to find a better-behaved (finite) limit of the
index where most of the theory decouples and only chiral $U(1)_R$
sector operators contribute. It is crucial that the exotic chiral
operators contribute as well, since we will use their putative
contributions to argue that they must be absent in theories of type C
and constrained in theories of type B. This limit will also
immediately shed light on the order of the divergence in
\Indexabelian\ and the nature of the divergent contribution in
\singleFVi. 

To that end, we consider the limit described in \GaddeUV\foot{There this limit was written as \eqn\Uirlimita{
p,q,t\to 0~, \ \ \ {\rm with}\ \ \Big\{z={pq\over t}, 
w=\sqrt{p\over q}\Big\} \ \to \ {\rm fixed}~.
}}
\eqn\Uirlimit{
|p|~, |q|~, |t|\ll|z|~, |w|~, \ \ \ |w|=1~,|z|\le1~, \ \ \ w=\sqrt{p\over q}~, \ \ \ z={pq\over t}~,
}
where we neglect corrections of order $\CO(|p|, |q|, |t|)$ to the various quantities we study. Here, $z$ and $w$ correspond to $U(1)_R$ and $j_1$ fugacities
respectively. In this limit, we have
\eqn\limittwo{
\CI_{U(1)_R}:=\Tr_{\CH}(-1)^F e^{-\beta \tilde\delta_{2\dot -}} z^{-(R+r)}w^{2j_1}~,
}
with contributions only from states
satisfying $E=-r$ and $j_2=-R$. As we will see below, there are no
$SU(2)_R$-charged states contributing and so we can take
$z$ to be a $U(1)_R$ fugacity.

We will refer to this limit as the \lq\lq chiral $U(1)_R$" limit of the index. For a general SCFT we have
\eqn\indexcompi{\eqalign{ &\CI^{\rm s.l.}_{U(1)_R\ \CC_{R,r(j_1,j_2)}}=0~,\cr&\CI^{\rm s.l.}_{U(1)_R\ \hat\CC_{R(j_1,j_2)}}=0~,\cr&\CI^{\rm s.l.}_{U(1)_R \ \bar\CE_{r(j_1,0)}}=(-1)^{2j_1}z^{-r}\chi_{j_1}\left(w\right)~,\cr&\CI^{\rm s.l.}_{U(1)_R\ \bar\CD_{0(j_1,0)}}=(-1)^{2j_1}z^{j_1+1}\chi_{j_1}\left(w\right)~,\cr&\CI^{\rm s.l.}_{U(1)_R\ \CD_{0(0,j_2)}}=0~.  }}
In this case, \singleFVi\ becomes
\eqn\singleFVi{\eqalign{
&\CI_{U(1)_R \ V}^{\rm s.l.}(p,q,t)=z~,\cr&\CI^{\rm s.l.}_{U(1)_R\ {H\over2}}(p,q,t)=0~.
}}
As a result, the divergence in \Indexabelian\ is regulated and
\eqn\Indexabeliani{
\CI_{U(1)_R}=\prod_a \exp\left(\sum_{n=1}^{\infty}{1\over n}z^n\right)={1\over(1-z)^N}~.
}
In particular, taking the additional limit $z\to1$, once again opens
up the Coulomb branch. We see that the divergence in this limit
corresponds to a pole of order $N$ (the complex dimension of the
Coulomb branch, ${\rm dim}\CM_\CC$) due to contributions from the $N$ different $\Phi_a$ scalars.

For a general Lagrangian SCFT, we have
\eqn\Indexlag{
\CI_{U(1)_R}=\prod_a \exp\left(\sum_{n=1}^{\infty}{1\over n}z^{E_an}\right)=\prod_a{1\over1-z^{E_a}}~,
}
where the $E_a$ are the scaling dimensions of the Casimirs, $\CO_a=\Tr\Phi^k$, and $a=1,\ldots, N$ parameterizes the directions along the Coulomb branch. In deriving \Indexlag, we have used the relation $E_a=-r_a$ (see \chiralops), the fact that the Casimirs are good coordinates on $\CM_{\CC}$, and, more generally, the fact that there are no operator relations among the Casimirs. We see again that the order of the pole at $z=1$ in \Indexlag\ corresponds to the dimension of the Coulomb branch.

In a more general $\CN=2$ theory, the direct analogs of the $\CO_a$ operators (i.e., the non-nilpotent (scalar) generators of the $U(1)_R$ chiral ring) might obey various constraints (although we are not aware of any examples in the currently published literature; note that we are also not aware of any examples of nilpotent $U(1)_R$ chiral ring generators). Such constraints are implemented by setting independent holomorphic functions, $f_i(\CO_a)$, with a well-defined $U(1)_R$ charge to zero
\eqn\constraint{
f_i(\CO_a)=0~, \ \ \ r(f_i)=-E_i<0~.
}
When we write the right hand side of the first equation in
\constraint, we work modulo descendants. We can include such constraints via a meromorphic factor, $\tilde I(z)$
\eqn\coulombgen{
\CI_{U(1)_R}=\prod_a{1\over1-z^{E_a}}\cdot\tilde \CI(z)~.
}
If there are no additional relations among the $f_i(\CO_a)$, then one can capture the full effect of the constraints via a \lq\lq wrong-sign" contribution of the form
\eqn\constraintInd{
\tilde \CI(z)=\prod_i\exp\left(-\sum_{n=1}^{\infty}{1\over n} z^{E_i}\right)=\prod_i(1-z^{E_i})~.
}
In this case, each constraint contributes a zero to the index in the limit $z\to1$, and we interpret it as \lq\lq lifting" a direction of the moduli space. If the number of $\CO_a$, $N_a$, is larger than the number of $f_i$, $N_i$, then we find a pole of order $N_a-N_i$ as $z\to1$, and we interpret the order of this pole as the complex dimension of the $U(1)_R$ moduli space.\foot{As an aside, we observe that for type A theories, the limit of the index we are studying is equivalent to the Hilbert series of the Coulomb branch. Various examples of Coulomb branch Hilbert series have been studied in \HananyUC. Note that this relation between the index and the Hilbert series seems to be more robust than the relation between the Hall-Littlewood index and the Hilbert series of the Higgs branch (see e.g., \BenvenutiPQ), which only holds for special genus zero classes of type A theories \GaddeUV.}

If our theory has exotic chiral operators, we get contributions to the index of the form
\eqn\exoticcont{\eqalign{
&\CI_{U(1)_R, \ j_1}=\prod_{a=1}^{N_{j_1}}\prod_{j=-j_1}^{j_1}{1\over1-z^{E_a}w^{2j}}~, \ \ \ j_{1}\in{\bf Z}~, \cr&\CI_{U(1)_R, \ j'_1}=\prod_{a=1}^{N_{j'_1}}\prod_{j=-j'_1}^{j'_1}(1-z^{E_a}w^{2j})~, \ \ \ j'_{1}\in{{\bf Z}+{1\over2}}~,
}}
where the first line in \exoticcont\ comes from $N_{j_1}$ spin $j_1$ chiral bosons, and the second line in \exoticcont\ comes from $N_{j'_1}$ spin $j'_1$ exotic chiral fermions. We can implement higher-spin constraints via a meromorphic function, $\tilde\CI(z,w)$, and find the general result
\eqn\coulombgeni{
\CI_{U(1)_R}=\prod_I\CI_{U(1)_R, j_I}\cdot\tilde \CI(z,w)~.
}
Before proceeding, let us note that $\tilde \CI(z,w)$ must be such that we have
\eqn\coulombgenii{
\CI_{U(1)_R}=\sum_{B}z^{E_B}\chi_{j_B}(w)-\sum_Fz^{E_F}\chi_{j_F}(w)~,
}
where $B$ and $F$ run over the full set of bosonic and fermionic chiral operators that contribute to the index taking into account all chiral operator relations. Crucially for us below, it follows that in any $\CN=2$ SCFT
\eqn\condIndi{
\partial_w\left(\lim_{z\to1}\CI_{U(1)_R}\right)\ne0 \ \ \ \Leftrightarrow \ \ \ \exists \ \CO_{\alpha_1\cdot\cdot\cdot\alpha_{2j_1}} \ {\rm s.t.} \ j_1>0~.
}
In other words, the $z\to1$ limit of the chiral $U(1)_R$ index is a non-trivial function of $w$ if and only if we have exotic chiral operators.

Let us next switch to the path integral interpretation. The limit
\Uirlimit\ involved taking:
\eqn\secondlimit{
|p|~, |q|~, |t|\ll|z|~, |w|~, \ \ \ |w|=1~,|z|\le1~, \ \ \ w=\sqrt{p\over q}~, \ \ \ z={pq\over t}~.
}
We found that only vector multiplets contribute (finitely), while
the hypermultiplets gave zero for all fields without the need for mutual cancelations. In terms of chemical potentials, and specializing to real and positive $p$, $q$, $t$, \secondlimit\ can be recast
into
\eqn\secondchem{\eqalign{
 &  v_z= v_p + v_q - v_t\ll1 \;,\quad v_w=
  \half v_p-v_q\ll1\;,\cr
&{\rm with} \quad
  v_p\sim\CO(1)\quad{\rm and}\quad \beta\gg1\;.   
}}
In order to proceed, consider
\eqn\alphalimit{\eqalign{
\alpha & =\Big(-2 j_2 - 2R + r
+ v_p(j_1 + j_2 - r ) + v_q (j_2 - j_1 - r) + v_t (R + r)\Big)\cr
  &= \Big( - 2 j_2 -2 R + r + v_z(j_2 - r ) +2 v_w
  j_1 + v_t (R + j_2)\Big)
}}
and focus once again on the masses for the constant modes. Using
\alphalimit\ one has
\eqn\masslimit{\eqalign{
M^2_s &
=(1-\alpha_s^2)\cr
&= (1- 2R + r - v_z r + v_t R ) (1 + 2R - r + v_z r - v_t R ) \;.
}}
Then for the vector multiplet and hypermultiplet scalars,
$(\Phi,Q)$, for which $r = -1$, $R = 0$ and $r = 0$, $R = \half$
respectively, we obtain
\eqn\scalarmasses{\eqalign{
M^2_{\Phi,s} & = v_z ( 2-v_z) \sim2v_z\ll1~,\cr
M^2_{Q,s} & =  v_t \left(1-{v_t\over 4}\right)\sim\CO(1)\;.
}}
Note that the vector multiplet scalars are parametrically lighter than the hypermultiplets. When we set $v_z=0$ to get $z=1$, we see that $M^2_{\Phi,s}=0$, and the Coulomb branch opens up.

For a theory in flat space, one can integrate down to an arbitrary momentum shell and obtain a two-derivative effective Lagrangian. However, for a theory on curved space one can never integrate out states at scales lower than the inverse radius of curvature and remain with an effective two-derivative theory.  The reason is that there is no way to separate higher derivative terms in the effective Lagrangian from terms suppressed by the inverse radius of curvature \FestucciaWS. This picture self-consistently reverts to the flat space intuition when the volume---and hence the radius of curvature---goes to infinity.

We will hence compare the masses \scalarmasses\ against the curvature,
which for $S^1 \times S^3$---and in units of the $S^3$ radius---simply
reads $ \CR \propto 1$. For some fixed $0<v_t<4$ this leads to a
hierarchy of scales:\foot{We have chosen the value of $v_t$ so that it
leads to a sensible field theory with real hypermultiplet
masses. Having said that, the index is well defined for all $v_t>0$.}
\eqn\hierarchy{\eqalign{
M^2_{\Phi,s}\ll \CR\sim M^2_{Q,s}  \;.
}}
As a result, in this second limit of chemical potentials, the
hypermultiplets can be integrated out and the index will once  again only
receive contributions from the vector multiplets. Since the index only
depends on the parameter $\beta$ (in units of the $S^3$ radius) and
chemical potentials, and in particular is independent of all other
continuous parameters including the RG scale \FestucciaWS, we can
conclude that there will be no contributions from the hypermultiplets
at any energy.

It may be useful to point out that a rescaling of both the $S^1$ and
$S^3$ radii by $\kappa$, which does not affect the index since their
ratio is kept fixed, will not disrupt the scale hierarchy. Although
the curvature changes as $\CR \to \kappa^{-2}\CR $, the
masses \scalarmasses\ are defined in units of the $S^3$ radius and
will also scale by $M^2\to \kappa^{-2} M^2$ leaving \hierarchy\
unchanged.

\subsec{Theories of Type B}

We now have the necessary tools to analyze theories of type B. Let us begin by discussing $SU(N)$ $\CN=2$ superconformal quantum chromodynamics (SCQCD) in more detail. Although we can immediately conclude from the preceding discussion that this theory does not have exotic chiral operators, a careful analysis of $SU(N)$ SCQCD and its Coulomb branch furnishes some important intuition about the behavior of the index in more general theories. From \Indexlag\ we have
\eqn\Indexlagii{
\CI_{U(1)_R}=\prod_{a=2}^N{1\over1-z^a}~.
}
In the limit $z\to1$, we find bosonic zero modes corresponding to the opening up of a Coulomb branch, and
\eqn\Indexlagiii{\eqalign{
\lim_{z\to1}\CI_{U(1)_R}=&|W_{SU(N)}|^{-1}(1-z)^{-{\rm rank}(SU(N))}+\CO\left((1-z)^{-{\rm rank}(SU(N))+1}\right)~,
}}
where $|W_{SU(N)}|=N!$ is the order of the Weyl group of $SU(N)$, and
${\rm rank}(SU(N))=N-1$ is the rank of the gauge group.

To understand this result, recall that the Coulomb branch of the $SU(N)$ theory is $\CM_{\CC}={{\bf C}^{N-1}/W_{SU(N)}}$. 
Furthermore, the $z\to1$ limit of the $SU(N)$ partition function on $M_4\simeq S^3\times S^1$ is dominated by field configurations with large vevs for the bosonic zero modes\foot{Similar reasoning was used in \GaiottoXA\ to study the effect of hypermultiplet zero modes on the superconformal index.}---precisely the regime where the theory should behave like a $U(1)^{N-1}$ theory on $\CM_C$. We divide by the order of the Weyl group since we should only integrate over the vector multiplet scalars modulo the Weyl identification. Therefore, \Indexlagiii\ conforms to our expectations: the $z\to1$ limits of the $SU(N)$ partition function and index are $1/N!$ times the $z\to1$ limits of the corresponding quantities for a $U(1)^{N-1}$ theory.

It is also interesting to consider the generalization of the
$|W_{SU(N)}|^{-1}$ factor to non-Lagrangian theories (where an
interpretation in terms of the order of the Weyl group is no longer
possible): it is precisely the geometrical factor we get from
restricting the partition function to field configurations that
parameterize $\CM_{\CC}$. For example, consider the following rank 1
non-Lagrangian theories: the AD theories discussed in \ArgyresJJ\
derived from the $N_f=1,2,3$ $SU(2)$ gauge theory---let us label these
theories ${\rm AD}_{1,2,3}$---and the $E_{6,7,8}$ theories of Minahan
and Nemeschansky \refs{\MinahanFG, \MinahanCJ}---let us label these
theories as ${\rm MN}_{6,7,8}$. The corresponding complex one-dimensional
Coulomb branches, $\CM_{\CC}^{{\rm AD}_i}$ and $\CM_{\CC}^{{\rm
MN}_i}$, are just complex cones with opening angles (in units of
$2\pi$) of $5/6, 3/4, 2/3$ and $1/3, 1/4, 1/6$ respectively (they are
part of a classification due to Kodaira \refs{\Kodaira}). 
\midinsert
\smallskip
\begintable
           $\CN = 2$ Theory| $ \vert W\vert^{-1} $ \crthick
           $SU(N)$ | $1/N!$\cr
           ${\rm AD}_1$ | 5/6\cr
           ${\rm AD}_2$ | 3/4\cr
           ${\rm AD}_3$ | 2/3\cr
           ${\rm MN}_6$ | 1/3\cr
           ${\rm MN}_7$ | 1/4\cr           
           ${\rm MN}_8$ | 1/6
\endtable
\Table{1}{Values for the leading-divergence index coefficient.}
\endinsert
The values of $|W|^{-1}$ required to appropriately restrict the
partition function are given in Table~1. These are precisely the
inverse dimensions of the Coulomb branch operators in these theories,
$E^{-1}$, and so they can also be understood as arising from the
$z\to1$ limit of $\CI_{U(1)_R}\supset {1\over1-z^{E}}\sim E^{-1}{1\over1-z}$.\foot{This
discussion also applies to the new rank one theories in
\refs{\ArgyresTQ}, since these latter theories have the same Coulomb
branch spectrum as the corresponding MN theories.} One interesting
check of this story is to note that the Coulomb branch of $SU(2)$
SCQCD is given by the cone ${\bf C}/W_{SU(2)}\simeq{\bf
C}/{\bf Z}_2$ and that therefore we should have $|W_{SU(2)}|^{-1}=1/2$
by our reasoning. Indeed, this result agrees with \Indexlagii\ and
\Indexlagiii.

We see that this argument is quite general and so we expect that for any theory, $\CT$, of type B there is a pole as $z\to1$ of order the complex dimension of the Coulomb branch multiplying a term describing the geometry on $\CM_{\CC}$.\foot{If the Coulomb branch consists of multiple sub-branches---potentially of different dimensions---then we expect one such term for each sub-branch.} Since the theories away from the origin of the Coulomb branch are Lagrangian (with the possible exception of theories living on certain complex co-dimension one or higher sub-manifolds of $\CM_{\CC}$), this term should be independent of $w$.  
Furthermore, we do not expect divergent contributions to the index coming purely from exotic operators since there is no supersymmetric moduli space associated with them.\foot{As we will see below, it is possible to relax this assumption at the cost of making some of our arguments more complicated.}

In other words, we should have that
\eqn\coulomblim{
\lim_{z\to1}\CI_{U(1)_R}(\CT)(z,w)=\lim_{z\to1}\left({\kappa\over(1-z)^{d_{\CC}}}+\cdot\cdot\cdot\right)~,
}
where $d_{\CC}={\rm dim}\CM_{\CC}(\CT)$, $\kappa$ is a constant, and the leading term on the right hand side describes the Lagrangian bulk of the Coulomb branch (if the theory has more than one Coulomb branch, then we have a similar term for each branch). The ellipsis in \coulomblim\ contains various lower-order terms (including any contributions in $w$).

One immediate consequence of \coulomblim\ is that if a theory of type B has exotic chiral operators, then for each exotic chiral operator, $\CO_{A\alpha_1\cdot\cdot\cdot\alpha_{2j_A}}$
, there exist non-negative integers, $N_{A, a,b}$, and dimensionless constants $c_b$ 
such that
\eqn\oprelats{
\CO_{A\alpha_1\cdot\cdot\cdot\alpha_{2j_A}}\cdot\sum_b\left(c_b\prod_a\CO_{a}^{N_{A,a,b}}\right)=0~, \ \ \ \sum_b\left(c_b\prod_a\CO_{a}^{N_{A,a,b}}\right)\ne0~,
}
where the $\CO_a$ are Coulomb branch operators (more precisely, we mean the non-nilpotent scalar generators of the $U(1)_R$ chiral ring). As in \constraint, we work modulo descendants (if there are multiple Coulomb branches, then we have an equation of the form \oprelats\ for each corresponding set of Coulomb branch operators).

Let us now briefly discuss the lower-order divergences in the index as $z\to1$. A priori, it is possible that sub-leading divergences in $1-z$ are $w$-dependent since there can be co-dimension one or higher sub-manifolds of $\CM_{\CC}$ that correspond to non-Lagrangian (or, more generally, to non-type A) theories. In the next section, however, we will rule out any $w$-dependence in the superconformal indices of theories that can flow to type A theories by turning on a relevant $\CN=2$-preserving coupling.

From the above discussion, it is not too difficult to see that the superconformal index of a type B theory can be written as follows 
\eqn\superconfdecomp{\eqalign{
\CI_{U(1)_R}=&\sum_{a_1, \cdot\cdot\cdot, a_{d_{\CC}}}{1\over1-z^{E_{a_1}}}\cdot\cdot\cdot{1\over1-z^{E_{a_{d_{\CC}}}}}\cdot\tilde\CI_{a_1\cdot\cdot\cdot a_{d_{\CC}}}(z)+\cr&\sum_{a_1, \cdot\cdot\cdot, a_{d_{\CC}-1}}{1\over1-z^{E_{a_1}}}\cdot\cdot\cdot{1\over1-z^{E_{a_{d_{\CC}-1}}}}\cdot\tilde\CI_{a_1\cdot\cdot\cdot a_{d_{\CC}-1}}(z,w)+\cr&\cdot\cdot\cdot+\sum_{a_1}{1\over1-z^{E_{a_1}}}\tilde\CI_{a_1}(z,w)+\tilde\CI(z,w)~,
}}
where $a_i\in\left\{1,\cdot\cdot\cdot, d_{\CC}, d_{\CC}+1,\cdot\cdot\cdot, d_{\CC}+n\right\}$ runs over the Coulomb branch generators, $\CO_{a_i}$, of dimension $E_{a_i}$ (let us note again that we mean the non-nilpotent scalar generators of the $U(1)_R$ chiral ring---more precisely, a minimal subset of these generators that give all the infinite-order contributions to the $U(1)_R$ chiral ring). If there are Coulomb branches of various dimensions, then $d_{\CC}$ refers to the maximal dimensional Coulomb branch(es). Note that in type A theories, $n=0$ and $a_i\in\left\{1, \cdot\cdot\cdot, d_{\CC}\right\}$.

In writing \superconfdecomp, we have factored out the infinite-order contributions from the Coulomb branch operators, and so the $\tilde\CI$ are non-singular. Furthermore, we take the $\tilde\CI$ to be symmetric in their indices, and if two indices take on the same value, then the corresponding $\tilde\CI$ vanishes (so as not to over-count the contributions from Coulomb branch operators).

For completeness, let us examine the $\tilde\CI$ in more detail. We define the $\tilde\CI$ so that the chiral operators contributing to $\tilde\CI_{a_1\cdots a_{d_{\CC}}}(z)$ satisfy $\tilde\CO\cdot\CO\ne0$ for any $\CO\ne0$ in the ring generated by $\left\{\CO_{a_1}, \cdot\cdot\cdot, \CO_{a_{d_{\CC}}}\right\}$. It therefore follows from \oprelats\ that $\tilde\CI_{a_1\cdots a_{d_{\CC}}}(z)$ is independent of $w$. There is, however, some ambiguity in the $z$-dependent contributions to $\tilde\CI_{a_1\cdots a_{d_{\CC}}}(z)$, since we may, for example, decide to include the contribution from $\tilde\CO$ itself in some lower-order $\tilde\CI$ function, or---if $\tilde\CO\cdot\CO'\ne0$ for any $\CO'\ne0$ in the ring generated by some other generators $\left\{\CO_{a_1'},\cdots\CO_{a'_{d_{\CC}}}\right\}$---in $\tilde\CI_{a_1'\cdots a_{d_{\CC}}'}(z)$.

If we include the contribution of $\tilde\CO$ in $\tilde\CI_{a_1\cdots a_{d_{\CC}}}(z)$, then we find a corresponding term of the form 
\eqn\cont{
{1\over d_{\CC}!}z^{\tilde E}\subset\tilde\CI_{a_1\cdots a_{d_{\CC}}}(z)~,
}
where $\tilde E$ is the scaling dimension of $\tilde\CO$ (and the factor $(d_{\CC}!)^{-1}$ comes from the symmetrization in the $a_i$ indices). On the other hand, if we choose to include the contribution from $\tilde\CO$ in some other $\tilde\CI$ function, then we should replace \cont\ with 
\eqn\contii{\eqalign{
&{1\over d_{\CC}!}z^{\tilde E}\cdot\Big(\sum_{a_i}z^{E_{a_i}}-\sum_{(a_{i_1},a_{i_2})}z^{E_{a_{i_1}}+E_{a_{i_2}}}+\sum_{(a_{i_1}, a_{i_2}, a_{i_3})}z^{E_{a_{i_1}}+E_{a_{i_2}}+E_{a_{i_3}}}-\cdots\cr&+(-1)^{d_{\CC}+1}z^{\sum_{a_i}E_{a_i}}\Big)\subset\tilde\CI_{a_1\cdots a_{d_{\CC}}}(z)~,
}}
where $(a_{i_1},\cdots, a_{i_k})$ takes values in the ${d_{\CC}!\over(d_{\CC}-k)!k!}$ unordered $k$-tuples built out of $\left\{a_1,\cdots, a_{d_{\CC}}\right\}$.

It follows from our discussion that $\tilde\CI_{a_1\cdots
a_{d_{\CC}}}(z)$ is independent of the superconformal $U(1)_R$ charge
when we set $z=1$ (the only dependence of \cont\ and \contii\ on the
superconformal $U(1)_R$ is through the scaling dimension of the chiral
operators). Furthermore, $\tilde\CI_{a_1\cdots a_{d_{\CC}}}(1)$ is
independent of whether we choose to include the contribution of
$\tilde\CO$ in $\tilde\CI_{a_1\cdots a_{d_{\CC}}}(1)$ or not---i.e.,
the left hand sides of \cont\ and \contii\ are both equal to $+1$ when
evaluated at $z=1$ (this matching is guaranteed by the fact that
removing a finite number of operator contributions cannot change the
leading divergence of the index).\foot{More generally, we may choose to
remove an infinite (but sub-leading) number of contributions in
$\tilde\CO$ from $\tilde\CI_{a_1\cdots a_{d_{\CC}}}(z)$. For example,
we can remove all contributions that are not of the form
$\tilde\CO\CO_{a_1}\CO$ by replacing \cont\ with $z^{\tilde
E+E_{a_1}}\subset\tilde\CI_{a_1\cdots a_{d_{\CC}}}$. Note that again
$\tilde\CI_{a_1\cdots a_{d_{\CC}}}(z)$ is independent of this
choice.} In other words, we see that the various $\tilde\CI$
functions that multiply the leading divergence in \superconfdecomp\
are well-defined (finite and independent of the ambiguities discussed
above)\foot{We have made the physical assumption that the divergences of the index / partition function arise from the existence of moduli spaces of vacua. Therefore, we have assumed that the theory does not contain an infinite number of chiral operators
that do not form a ring, as there is no moduli space associated with them. If the theory has an infinite number of Coulomb branches, the
index will of course not be finite, although the origin of the
divergence is different from the one we are considering here.} and universal (independent of the superconformal $U(1)_R$) at $z=1$, with
\eqn\Izone{
d_{\CC}!\cdot \tilde\CI_{a_1\cdots a_{d_{\CC}}}(1)=\sum_B1=n_B > 0~.
}
Note that the general form of $\tilde\CI_{a_1\cdots a_{d_{\CC}}}(1)$, as a sum over positive contributions from contributing operators, is independent of which minimal subset of generators we choose to use in writing the infinite order contributions in \superconfdecomp.

Next, let us proceed to the order $d_{\CC}-1$ terms (i.e., the $\tilde\CI$ in the second sum in \superconfdecomp). These terms may a priori receive contributions from exotic chiral operators. Since we have factored out the infinite-order contributions from the Coulomb branch operators, the $\tilde\CI_{a_1\cdots a_{d_{\CC}-1}}(z,w)$ are also finite. Furthermore, these functions are independent of the superconformal $U(1)_R$ charge when we set $z=1$. Finally, in analogy to the discussion above, $\tilde\CI_{a_1\cdots a_{d_{\CC}-1}}(1,w)$ is independent of whether we choose to include the contribution of a particular exotic operator, $\tilde\CO_{\alpha_1\cdots\alpha_{2j}}$,  (or even an infinite but subleading number of similar contributions from other exotic operators) in it or not since we either have
\eqn\contex{
{(-1)^{2j}\over (d_{\CC}-1)!}z^{\tilde E}\chi_{j}(w)\subset\tilde\CI_{a_1\cdots a_{d_{\CC}-1}}(z,w)~,
}
or
\eqn\contexii{\eqalign{
&{(-1)^{2j}\over(d_{\CC}-1)!}z^{\tilde E}\chi_j(w)\cdot\Big(\sum_{a_i}z^{E_{a_i}}-\sum_{(a_{i_1},a_{i_2})}z^{E_{a_{i_1}}+E_{a_{i_2}}}+\sum_{(a_{i_1}, a_{i_2}, a_{i_3})}z^{E_{a_{i_1}}+E_{a_{i_2}}+E_{a_{i_3}}}-\cdots\cr&+(-1)^{d_{\CC}+1}z^{\sum_{a_i}E_{a_i}}\Big)\subset\tilde\CI_{a_1\cdots a_{d_{\CC}-1}}(z,w)~.
}}
Clearly, \contex\ and \contexii\ are both equal to $\chi_j(w)$ at $z=1$. Therefore, we find that the exotic operator contributions to $\tilde\CI_{a_1\cdots a_{d_{\CC}-1}}(1,w)$ are unambiguously given by
\eqn\Izoneex{
(d_{\CC}-1)!\cdot\tilde\CI_{a_1\cdots a_{d_{\CC}-1}}(1,w)|_{j\ne0}=\sum_{j\in{\bf Z}>0}n_{(j)a_1\cdots a_{d_{\CC}}-1}\cdot\chi_j(w)-\sum_{j\in{{\bf Z}+{1\over2}>0}}n_{(j)a_1\cdots a_{d_{\CC}}-1}\cdot\chi_j(w)~,
}
where the $n_{(j)}$ are non-negative integers.

In particular, it is clear that the order $d_{\CC}-1$ contribution to the index depends on $w$ if and only if this is true in the limit $z\to1$. Furthermore, if the order $d_{\CC}-1$ contribution is independent of $w$, then the order $d_{\CC}-2$ contribution is independent of $w$ if and only if this is true in the limit $z\to1$ (and so on sequentially down to the finite contributions). Finally, let us note that the divergences multiplying the various $\tilde\CI$ depend on the superconformal $U(1)_R$ only through a positive overall factor (times $(1-z)^{-n}\tilde\CI$) given by the inverse product of (minus) the $U(1)_R$ charges of the corresponding Coulomb branch operators.

\subsec{Theories of Type C}
In this subsection, we wish to discuss theories that flow to type A theories when we turn on a deformation by a relevant coupling. Since we want to preserve $\CN=2$ invariance, we can turn on relevant deformations of the prepotential
\eqn\reldefpre{
\delta\CL=\int d^4\theta\CO+{\rm h.c.}~,
}
where $d^4\theta$ is the $\CN=2$ chiral integration measure, and $\CO$ is a generator of the chiral ring with dimension $E(\CO)<2$. In addition, we can consider deformations of the superpotential by a moment map, $\mu$, corresponding to some symmetry, $G$, of the UV SCFT
\eqn\reldefW{
\delta\CL=\int d^2\theta\mu+{\rm h.c.}~.
}
As far as we are aware, all known $\CN=2$ SCFTs admit deformations by relevant operators that preserve $\CN=2$ SUSY (for example, while the $T_N$ theories do not have relevant deformations of the type described in \reldefpre, we can deform their superpotentials as in \reldefW). Note that we will not consider deformations by FI terms (in any case, the corresponding $U(1)$ gauge factors are not UV-complete).

Both types of relevant deformations in \reldefpre\ and \reldefW\ break the $U(1)_R$ symmetry of the UV SCFT (but preserve the $SU(2)_R$ symmetry). Therefore, if we wish to compute the supersymmetric index of the interpolating theory, we should turn off the fugacity corresponding to $U(1)_R$, i.e., we should take $z\to1$. It is again useful to study the $U(1)_R$ limit of the index. In the UV we find that the index is
\eqn\superconfdecompi{\eqalign{
\CI_{U(1)_R}=&\sum_{a_1, \cdot\cdot\cdot, a_{d_{\CC}}}{1\over1-z^{E_{a_1}}}\cdot\cdot\cdot{1\over1-z^{E_{a_{d_{\CC}}}}}\cdot\tilde\CI_{a_1\cdot\cdot\cdot a_{d_{\CC}}}(z,w)+\cr&\sum_{a_1, \cdot\cdot\cdot, a_{d_{\CC}-1}}{1\over1-z^{E_{a_1}}}\cdot\cdot\cdot{1\over1-z^{E_{a_{d_{\CC}-1}}}}\cdot\tilde\CI_{a_1\cdot\cdot\cdot a_{d_{\CC}-1}}(z,w)+\cr&\cdot\cdot\cdot+\sum_{a_1}{1\over1-z^{E_{a_1}}}\tilde\CI_{a_1}(z,w)+\tilde\CI(z,w)~.
}}
This formula for the index is very similar to the one in \superconfdecomp, but we allow for the leading term to be $w$-dependent since we do not assume that the $U(1)_R$ moduli space, $\CM_{U(1)_R}$, parameterized by the vevs of the (non-nilpotent) chiral generators, $\CO_{a_i}$, is a Coulomb branch (i.e., we allow for a fully interacting theory along the whole moduli space, although we do not know of any such examples). In this case, by reasoning similar to that used in the previous subsection, the leading $\tilde\CI$ functions have the following general form at $z=1$
\eqn\leadingCIex{
d_{\CC}!\cdot\tilde\CI_{a_1\cdots a_{d_{\CC}}}(1,w)=\sum_{j\in{\bf Z}\ge0}n_{{(j)a_1\cdots a_{d_{\CC}}}}\cdot\chi_j(w)-\sum_{j\in {\bf Z}+{1\over2}>0}n_{{(j)a_1\cdots a_{d_{\CC}}}}\cdot\chi_j(w)~,
}
where the $n_{(j)}$ are non-negative integers.

As we discussed in the previous subsection, the index in \superconfdecompi\ will generally diverge as we take $z\to1$. However, as we mentioned above, the $\tilde\CI$ functions are finite and do not depend on the (broken) superconformal $U(1)_R$ symmetry when we set $z=1$. We can therefore associate a universal meaning to \leadingCIex\ at all length scales, including in the deep IR. A useful way to extract the $\tilde\CI$ is to consider the following contour integrals around $z=1$:
\eqn\wint{
\CA_{n+1}(w)=\oint{dz\over2\pi i}(1-z)^n\cdot\CI_{U(1)_R}(z,w)~.
}
We can then pick out $w$-dependent quantities in the $\CA_{n+1}$ by defining
\eqn\winta{
\CB_{j_i, n+1}=\oint{dw\over2\pi iw}\Delta(w)\chi_{j_i}(w)\CA_{n+1}(w)~, \ \ \ \Delta(w)={1\over2}(1-w^2)(1-w^{-2})~.
}

Let us now compute the various $\CA_{n+1}$ in the UV and compare them to the corresponding quantities in the IR. We begin with $\CA_{d_{\CC}}$ (for $n+1>d_{\CC}$, the $\CA_{n+1}$ vanish)
\eqn\Amax{
\CA_{d_{\CC}}^{UV}(w)=-\sum_{a_1,\cdot\cdot\cdot, a_{d_{\CC}}}\left(\prod_{i} E_{a_i}^{-1}\cdot\tilde\CI_{a_1\cdot\cdot\cdot a_{d_{\CC}}}(1,w)\right)~.
}
The superconformal $U(1)_R$ dependence of the above expression is
contained in the inverse UV scaling dimensions of the $\CO_{a_i}$. In
the IR we can then compute the same quantity\foot{Note that the number
of non-nilpotent $\CO_{a_i}$ generators do not change along our
class of RG flows (although their dimensions will generally change,
and operators parameterizing the same branch will generally mix). The total $n_{(j)}$ parameterizing a given branch are also invariant.} and find
\eqn\AmaxIR{
\CA_{d_{\CC}}^{IR}(w)=-\sum_{\hat a_1,\cdot\cdot\cdot, \hat a_{d_{\CC}}}\left(\prod_{i} \hat E_{\hat a_i}^{-1}\cdot\tilde\CI_{\hat a_1\cdot\cdot\cdot \hat a_{d_{\CC}}}(1,w)\right)~,
}
where $\hat E_{\hat a_i}$ are the (generally different) scaling
dimensions of the IR $\hat\CO_{\hat a_i}$ operators. While it is usually true that $\CA_{d_{\CC}}^{UV}(w)\ne \CA_{d_{\CC}}^{IR}(w)$, it is also clear that since the $E_{a_i}, \hat E_{\hat a_i}>0$ and the $\tilde\CI_{a_1\cdot\cdot\cdot a_{d_{\CC}}}$ have the form given in \leadingCIex, then $\CA^{UV}_{d_{\CC}}$ is $w$-dependent if and only if $\CA^{IR}_{d_{\CC}}$ is $w$-dependent. Said differently, $\CB^{UV}_{j_i, d_{\CC}}\ne0$ if and only if $\CB^{IR}_{j_i, d_{\CC}}\ne0$.

Suppose now that the IR theory is type A. In this case, we know that $\CB^{IR}_{j_i, d_{\CC}}=0$ for all $j_i\ne0$. Therefore, the same must hold in the UV. In other words, \leadingCIex\ is independent of $w$. Proceeding sequentially down the list of $\CB^{UV, IR}_{j_i, n+1}$, we find that if the IR theory is of type A, then the UV theory has $\CB^{UV}_{j_i, n+1}=0$ for all $n$. In other words, the UV theory cannot have exotic operators.\foot{In the discussion so far, we have assumed that the exotic chiral operators give finite contributions to the superconformal index (since they are not associated with a moduli space). If we relax this assumption, we can repeat the discussion around \superconfdecomp\ and \superconfdecompi\ with the corresponding additional poles factored out.}

\subsec{Examples of type C theories}
Let us now consider some particular cases of non-Lagrangian type C
theories. For example, we can consider the $N_f=2$ AD theory discussed
in \ArgyresJJ. This theory has a single relevant prepotential coupling. It corresponds to an operator, $\CO$, of
dimension $4/3$. At the superconformal point, the Seiberg--Witten curve is
\eqn\curveNfii{
x^2=z^4~.
}
Deforming the theory by $\delta\CL\sim\int d^4\theta\ u_2\CO+{\rm h.c.}$ is equivalent to deforming \curveNfii\ as follows
\eqn\curveNfiii{
x^2=z^4+u_2z^2~.
}
In the deep IR we recover (in addition to a decoupled $U(1)$)
\eqn\curveIR{
x^2=z^2~,
}
which is the curve of a free Lagrangian theory and is therefore of type A.

As a result, we conclude that the $N_f=2$ AD theory does not contain
exotic chiral operators. More generally, we can take any of the
infinitely many $(G,G')$ AD SCFTs for $G,G' = A_n,D_n,E_n$
\refs{\ShapereXR, \CecottiFI}\ and sequentially deform them to
Lagrangian theories via relevant deformations (the theory in
\curveNfii\ is the $(A_1, A_3)$ theory). We explain this in some
detail in Appendix B. From our above reasoning, we conclude that these
theories cannot have exotic chiral operators. It would be
interesting to understand the full set of theories that can be
deformed to type A ones---we suspect that this is a very large class.

\newsec{Conformal Anomalies and the Index}

As we discussed above, the superconformal index is a robust observable
in any SCFT, $\CT$, since it is invariant under exactly marginal
deformations. Moreover, the superconformal index captures global
information about the protected sector of the theory. Similarly, the
$a$ and $c$ conformal anomalies are invariant under exactly marginal
deformations and describe global information about the theory (since
they measure $\CT$'s response to a background metric). Therefore, it
is natural to imagine that there is some relation between the index
and the conformal anomalies (see also the recent holographic
discussion in \refs{\ArdehaliXYA,\ArdehaliXLA} and the more general
analysis in \DiPietroBCA).\foot{Note that unlike in the interesting
recent work \AsselPAA, here we have in mind a direct relation between
the index---not the $M_4\simeq S^3\times S^1$ partition function---and
the conformal anomalies. Note also that we are extracting information
from the full index, as opposed to the gauge integrand considered
e.g. in \SpiridonovWW, as this approach is useful even in the context
of non-Lagrangian theories.}

This idea can be made more concrete in an $\CN=2$ SCFT, $\CT$, with a freely generated Coulomb branch (i.e., a Coulomb branch with no relations between the generators; as far as we are aware, all known $\CN=2$ SCFTs are of this type) and a Seiberg-Witten description by recalling that \refs{\ArgyresTQ,\ShapereZF}
\eqn\twoamc{
2a-c={1\over2}\sum_a\left(E_a-{1\over2}\right)~,
}
where $a$ and $c$ are the anomaly coefficients of $\CT$, and the $E_a$ are the scaling dimensions of the $\bar\CE_{r(0,0)}$ operators. We have seen that the $\bar\CE_{r(0,0)}$ operators contribute to the chiral $U(1)_R$ index with a weight determined by their conformal dimension. Thus, it is clear that we should be able to extract the right hand side of \twoamc, and therefore $2a-c$, from the index.

In order to understand how to make contact with \twoamc\ using the
index, it turns out to be useful to recall how the authors of
\ShapereZF\ derived \twoamc\ by building on the work of \refs{\WittenGF,\MoorePC}. The main strategy of \ShapereZF\ was to topologically twist $\CT$ and couple it to a background metric.  Under these conditions, the anomalous shift in the $U(1)_R$-charge of the vacuum becomes\foot{Our conventions for the $\CN=2$ $U(1)_R$ charge differ from those in \ShapereZF. In particular, we have $r_{\rm here}=-{1\over2}r_{\rm there}$ while keeping our conventions for $a$ and $c$ as well as the $SU(2)_R$ and $\CN=1$ $U(1)_R$ symmetries the same as in \ShapereZF.}
\eqn\Rchargeshift{
\Delta r=-(2a-c)\chi-{3\over2}c\sigma+{1\over2}k_Gn~,
}
where $\chi$, $\sigma$, $n$, and $k_G$ correspond to the Euler character, the
signature, the instanton number of the background, and the
flavor anomaly respectively. Since we are interested in $2a-c$, we will focus on the term in \Rchargeshift\ proportional to $\chi$.

In an $\CN=2$ SCFT with a Coulomb branch, $\CM_{\CC}(\CT)$, we can easily compute \Rchargeshift\ at a generic point on $\CM_{\CC}(\CT)$. Indeed, since the theory at such a point consists of $d_{\CC}={\rm dim}\CM_{\CC}(\CT)$ weakly coupled abelian vector multiplets, we simply have that
\eqn\twoacconta{
2a-c={d_{\CC}\over4}~.
}
On the other hand, when we approach the (strongly coupled) $\CN=2$ SCFT point it is more difficult to compute \Rchargeshift. However, it turns out that the correction to \twoacconta\ from the additional massless matter of the SCFT is encoded in the topologically twisted partition function, $Z$. 

To see this, note that $Z$ reduces to an integral over the (generically abelian) low energy effective theory
\eqn\topotwist{
Z=\int[du][dq]A^{\chi}B^{\sigma}C^ne^{-S_{\rm low \ energy}}~,
}
where $[du]$ and $[dq]$ correspond to the vector multiplets and hypermultiplets that parameterize the abelian moduli space (i.e., these fields are massless everywhere), and the holomorphic measure factors $A$, $B$, and $C$ give corrections to the low energy theory arising from matter that becomes massless only at special points on the moduli space. In particular, we have that $2a-c$ at the superconformal point is just
\eqn\twoaccontb{
2a-c=-r(A)+{d_{\CC}\over4}=E(A)+{d_{\CC}\over4}~,
}
where $r$ is the $U(1)_R$ charge of $A$ and $E$ is its scaling dimension. The beautiful observation of the authors in \refs{\WittenGF,\MoorePC,\ShapereZF} is that $A$ can be determined in terms of the data of the low energy theory (\refs{\MoorePC} was the first work to show this fact in the case of non-abelian gauge theory). In particular, $A$ is just the Jacobian
\eqn\Arelat{
A\sim\left(\det{\partial u_a\over\partial a^I}\right)^{1\over2}~,
}
where the $u_a$ are the gauge and monodromy-invariant coordinates and the $a^I$ are the special electric coordinates. The correction to $2a-c$ can then be computed since the dimensions of the $u_a$ are just the dimensions, $E_a$, of the $\bar\CE_{r(0,0)}$ multiplets of the SCFT (here we assume that the Coulomb branch is freely generated), while the $a^I$ have dimension one. In particular, we have
\eqn\dimjac{
E(A)={1\over2}\sum_a\left(E_a-1\right)~.
}
We then see that \dimjac\ together with \twoaccontb\ imply \twoamc.

Let us now see how to make contact with \twoamc\ using the
index. Surprisingly, even though our theory is not topologically
twisted, there is, as in \twoaccontb, a natural index decomposition of
the contributions to $2a-c$ from weakly coupled vector multiplets that
are massless in the bulk of $\CM_{\CC}$ and corrections from the
SCFT.\foot{Perhaps this is due to some relation between topological
twisting and the holomorphic twisting we employ in constructing the
index.} Just as in \ShapereZF, we will suppose that the Coulomb branch
is freely generated. For simplicity, let us also suppose that the
theory is of type A or type C so that we can drop any potential exotic contributions to the index (this result then applies to a broad class of theories including the $T_N$ theories and the AD theories with a Seiberg-Witten description; however, the expressions we will derive make sense for any type A or C $\CN=2$ theory). Under these assumptions, the general $U(1)_R$ index in \superconfdecomp\ becomes
\eqn\freegen{
\CI_{U(1)_R}=\prod_a{1\over1-z^{E_a}}~.
}
Since the derivation of \ShapereZF\ used properties of the theory on the Coulomb branch, it is natural to take the limit $z\to1$ in \freegen. In this limit, we find that
\eqn\limztoonefree{
\lim_{z\to1}\CI_{U(1)_R}=\prod_a{E_a^{-1}}\cdot(1-z)^{-d_{\CC}}+\CO\left((1-z)^{-d_{\CC}+1}\right)~,
}
where the number of degrees of freedom in the bulk of the Coulomb branch, $d_{\CC}$, can be read off from the leading divergence of the index---in particular, we see that we should include $d_{\CC}$ vector multiplets in computing the weakly coupled contribution to $2a-c$. This result is as we expect since the theory should behave to first approximation as the abelian theory on $\CM_{\CC}$.

Just as in \ShapereZF, we would like to see if it is possible to extract the SCFT corrections to $2a-c$. Therefore, it is natural to consider the leading corrections to \limztoonefree
\eqn\limztoonesl{
\lim_{z\to1}\CI_{U(1)_R}=\prod_a{E_a^{-1}}\left((1-z)^{-d_{\CC}}-E(A)\cdot (1-z)^{-d_{\CC}+1}\right)+\CO\left((1-z)^{-d_{\CC}+2}\right)~.
}
In particular, we see that the leading correction precisely captures the scaling dimension of the Jacobian \Arelat\ that describes the change in coordinates between the free $U(1)$ variables and vevs of SCFT operators.

We can then give a compact formula for $2a-c$ in terms of contour integrals of the index around the point $z=1$. Recalling the definitions in \wint\
\eqn\defnAn{
\CA_{n+1}=\oint {dz\over2\pi i} (1-z)^n\cdot\CI_{U(1)_R}(z)~,
}
it follows that
\eqn\dCEa{\eqalign{
&d_{\CC}=n_{\rm max}=\max\left\{n|\CA_{n}\ne0\right\}~,\cr& E(A)={\CA_{n_{\rm max}-1}\over\CA_{n_{\rm max}}}~.
}}
Therefore, we arrive at the result
\eqn\twoacind{
2a-c={\CA_{n_{\rm max}-1}\over\CA_{n_{\rm max}}}+{n_{\rm max}\over4}~.
}

As a final aside, let us note that our results also apply to the
subclass of the above theories with good holographic duals, where
$2a-c\sim\CO(N^2)\gg1$. Although  the index itself scales as
$\CO(N^0)$, and the residues we compute in \defnAn\ are heavily suppressed in the $1/N$ expansion (note, however, that the ratio of residues in \twoacind\ is not suppressed), we can read off the conformal anomaly by varying the
fugacities so that the index approaches a pole in the $z$ plane. The precise way in which the index behaves in the vicinity of the pole encodes the conformal anomalies (at least one linear combination of them).

\newsec{Summary and Conclusions}

In this paper we have derived constraints on higher-spin chiral
operators in large classes of $\CN = 2$ SCFTs. The first family, which
we denoted ``type A'', consisted of theories related to Lagrangian
SCFTs by generalized Argyres--Gaiotto--Seiberg duality. We used this
duality in combination with general symmetry arguments to argue that
type A theories do not contain the above ``exotic'' chiral operators
in their spectrum.

We then considered a ``chiral $U(1)_R$'' limit of the $\CN = 2$
superconformal index, which only receives contributions from
Coulomb branch operators and their cousins, as a tool towards investigating theories of
``type B''---generic theories with a Coulomb branch---and theories of
``type C'', which flow to type A via a deformation by a relevant coupling. In doing
so, we elaborated on properties of the chiral $U(1)_R$ index from the
perspective of its $S^1\times S^3$ path-integral interpretation. We
subsequently used this index to argue that theories of type C do not
possess exotic operators.

In the final section, we used the formula $2a -c = \half \sum_a (E_a -\half)$ of
\ShapereZF, for theories of type A or type C with a freely generated Coulomb
branch and a Seiberg-Witten description,\foot{This class of theories is very large and includes the $T_N$ theories as well as the infinite set of $(A_N, A_M)$ AD theories.} to give a prescription for reading off $2a - c$ directly from
the chiral $U(1)_R$ index. Our expressions were also valid for
examples of such theories with good holographic duals.

Note that our formula in \twoacind\ is actually well-defined for any type A or C $\CN=2$ SCFT (more generally, we are not aware of any concrete theory for which it is ill-defined). It would therefore be interesting to study whether our prescription for extracting $2a-c$ applies directly to
theories with a non-freely generated Coulomb branch (assuming such theories exist). If our prescription applies to this class of theories, then we can immediately constrain the possible set of Coulomb branch operator relations. For example, let us consider an abstract SCFT with Coulomb branch operators $\CO_{1,2}$ satisfying the constraint $\CO_1^M =\CO_2^N$. A direct application of our formula leads to a violation of the Hofman--Maldacena bound \HofmanAR\ unless $M=1$ or $N=1$---this reasoning would suggest that such constraints with $M, N\ne1$ should not be allowed in a unitary $\CN=2$ SCFT.

Finally, it is natural to ask whether, in direct analogy to the above
case of $2a-c$, one could use the index to also determine the $a-c$
linear combination of conformal anomaly coefficients (see also the
discussion in \DiPietroBCA). The latter can often be related to the
number of effective vector and hypermultiplets in the theory, which in
turn is proportional to the dimension of the Higgs branch (when it
exists), $a-c = {1\over24}(n_V- n_H) =-{1\over24}{\rm dim}\CH$. It would then be
reasonable to look for a limit of the superconformal index that also
counts Higgs-branch operators and attempt to extract $a-c$ along the
lines that we have discussed (although we should be careful to note
that one must deal with theories that don't have genuine Higgs
branches). This is another question that we will leave open as an
attractive direction for future research.

\vskip 1cm

\noindent {\bf Acknowledgments:}

We would like to thank Zohar Komargodski, Greg Moore, Leonardo
Rastelli, Nathan Seiberg, David Shih and Yuji Tachikawa for helpful
discussions and comments. This work was partly supported by the
U.S. Department of Energy under the grants DOE-SC0010008,
DOE-ARRA-SC0003883 and DOE-DE-SC0007897. CP is a Royal Society
Research Fellow.

\appendix{A}{$\CN= 2$ superconformal multiplet conventions}

In this appendix we provide a brief summary of our conventions for
quick reference. We follow Dolan and Osborn \DolanZH. A review of
their results\foot{These are based on earlier work of
\refs{\DobrevQV\DobrevVH-\DobrevQZ}, which established the
classification of unitary representations of $\frak{su}(2,2|N)$ for
all $N$, and studied their unitarity bounds and shortening
conditions.}  can be found in the appendices of \BeemSZA, to which we
point the interested reader for a concise account.

We work with the four-dimensional $\CN = 2$ superconformal algebra $\frak{s u}(2,2|2)$ in terms of a spinorial notation with $\alpha = +,-$  and
$\dot\alpha = \dot + , \dot - $. The purely
bosonic subalgebra can be captured by the following expressions:
\eqn\sca{\eqalign{
[{\CM_\alpha}^\beta, {\CM_\gamma}^\delta]  = \delta_\gamma^\beta
{\CM_\alpha}^\delta - \delta_\alpha^\delta {\CM_\gamma}^\beta~, \qquad&
[{\CM^{\dot\alpha}}_{\dot\beta}, {\CM^{\dot\gamma}}_{\dot\delta}]  =
\delta^{\dot\alpha}_{\dot \delta}
{\CM^{\dot\gamma}}_{\dot\beta} - \delta^{\dot\gamma}_{\dot\beta}
{\CM^{\dot\alpha}}_{\dot\delta} \cr
[{\CM_\alpha}^\beta, \CP_{\gamma\dot\gamma}]  = \delta_\gamma^\beta
\CP_{{\alpha\dot\gamma}} - \half\delta_\alpha^\beta
\CP_{\gamma\dot\gamma}~, \qquad& 
[{\CM^{\dot\alpha}}_{\dot\beta}, \CP_{\gamma\dot\gamma}]  =
\delta^{\dot\alpha}_{\dot \gamma}
\CP_{\gamma\dot\beta} - \half\delta^{\dot\alpha}_{\dot\beta}
\CP_{\gamma\dot\gamma}\cr
[{\CM_\alpha}^\beta, \CK^{\dot\gamma\gamma}]  = - \delta_\alpha^\gamma
\CK_{{\dot\gamma\beta}} + \half\delta_\alpha^\beta
\CK_{\dot\gamma\gamma}~, \qquad& 
[{\CM^{\dot\alpha}}_{\dot\beta}, \CK^{\dot\gamma\gamma}] = -
\delta^{\dot\gamma}_{\dot\beta}\CK^{\dot\alpha\gamma} + \half\delta^{\dot\alpha}_{\dot\beta}
\CK^{\dot\gamma\gamma}\cr
[\CK^{\dot\alpha\alpha},\CP_{\beta\dot\beta}] =
\delta_{\beta}^{\alpha}\delta^{\dot\alpha}_{\dot\beta} E+\delta_{\beta}^{\alpha}
\CM^{\dot\alpha}_{\dot\beta}+\delta^{\dot\alpha}_{\dot\beta}\CM_{\beta}^{\alpha}~,
\qquad & [{\CR^I}_J,{\CR^K}_L] =\delta^K_J
{\CR^I}_L-\delta^I_L{\CR^K}_J\cr
[E,\CP_{\alpha\dot\alpha}] =	\CP_{\alpha\dot\alpha}~, \qquad&
[E,\CK^{\dot\alpha\alpha}] =- \CK^{\dot\alpha\alpha}~,
}}
where 
\eqn\rottrans{
[{\CM_\alpha}^\beta] = \pmatrix{j_1 & j_1^+\cr j_1^- & - j_1}~,\qquad
[{\CM^{\dot\beta}}_{\dot\alpha}] = \pmatrix{j_2 & j_2^+\cr j_2^- & - j_2}
}
and
\eqn\Rmatrix{
[{\CR^I}_J]  = \pmatrix{R & R^+\cr R^- & -R\cr} +\half \pmatrix{r & 0
\cr 0& r\cr}\;,
}
such that 
\eqn\Rsymm{
[R^+, R^- ] = 2 R~, \qquad [R, R^\pm]  = \pm R^\pm\;.
}

The Poincar\'e and superconformal supercharges obey the relations
\eqn\susy{\eqalign{
\{\CQ_{\alpha}^I,\tilde\CQ_{J\dot\alpha}\} &=	\delta^I_J \CP_{\alpha\dot\alpha}\cr
\{\tilde\CS^{I\dot\alpha},\CS_J^{\alpha}\} &= \delta^I_J
\CK^{\dot\alpha \alpha}\cr
\{\CQ_{\alpha}^I,\CS^{\beta}_J\} &=\half \delta^I_J \delta_{\alpha}^{\beta}E   + \delta^I_J {\CM_{\alpha}}^{\beta}-\delta_\alpha^{\beta} {\CR^I}_{J}\cr
\{\tilde\CS^{I\dot\alpha},\tilde\CQ_{J\dot\beta}\} &=	\half \delta^I_J\delta^{\dot\alpha}_{\dot\beta}E + \delta^I_J {\CM^{\dot\alpha}}_{\dot\beta}+\delta^{\dot\alpha}_{\dot\beta} {\CR^I}_{J}~.
}}
These in turn satisfy the hermiticity conditions
\eqn\herm{
(\CQ^I_\alpha)^\dagger = \CS^\alpha_I~,\quad
(\tilde\CQ_{I\dot\alpha})^\dagger = \tilde\CS^{I \dot\alpha}~,\quad
({R^I}_J)^\dagger = {R^J}_I~,\quad ({\CM_\alpha}^\beta)^\dagger =
\delta^{\dot\beta\gamma}{\CM_\gamma}^\delta \delta_{\delta\dot\alpha}\;.
}

For the bosonic and fermionic generators we also have
\eqn\bosferm{\eqalign{
[{\CM_\alpha}^{\beta},\CQ_{\gamma}^I] =	\delta_{\gamma}^{\beta} \CQ_{\alpha}^I -\half\delta_{\alpha}^{\beta} \CQ_{\gamma}^I ~,
\qquad & [{\CM^{\dot\alpha}}_{\dot\beta},\tilde\CQ_{I \dot\delta}] =
\delta^{\dot\alpha}_{\dot\delta}\tilde\CQ_{I \dot\beta} -\half \delta^{\dot\alpha}_{\dot\beta}\tilde\CQ_{I \dot\delta}\cr
[{\CM_{\alpha}}^{\beta},\CS_{I}^{\gamma}] = -\delta_{\alpha}^{\gamma}\CS_{I}^{\beta}+\half\delta_{\alpha}^{\beta} \CS_{I}^{\gamma} ~,\qquad &
[{\CM^{\dot\alpha}}_{\dot\beta},\tilde\CS^{I\dot\gamma}] =
-\delta^{\dot\gamma}_{\dot\beta}\tilde\CS^{I\dot\alpha} + \half\delta^{\dot\alpha}_{\dot\beta}\tilde\CS^{I\dot\gamma}\cr
[E,\CQ_{\alpha}^I] = 	\half \CQ_{\alpha}^I ~,\qquad &
[E,\tilde\CQ_{I \dot\alpha}] = \half \tilde\CQ_{I \dot\alpha}\cr
[E, \CS_{I}^{\alpha}]	=-\half \CS_{I}^{\alpha}~,\qquad &
[E, \tilde\CS^{I\dot\alpha} ]	= -\half \tilde\CS^{I\dot\alpha}\cr
[{\CR^I}_J,\CQ_{\alpha}^K] = \delta_{J}^{K} \CQ_{\alpha}^I -{1\over 4} \delta_{J}^{I} \CQ_{\alpha}^\CK ~,\qquad &
[{\CR^I}_{J},\tilde\CQ_{K \dot\alpha}]	=-\delta_{K}^{I} \tilde\CQ_{J
\dot\alpha} +{1\over 4} \delta_{J}^{I} \tilde\CQ_{K \dot\alpha}\cr
[\CK^{\dot\alpha\alpha},\CQ_{\beta}^I] =	\delta_{\beta}^{\alpha}\tilde\CS^{I\dot\alpha}~,\qquad &
[\CK^{\dot\alpha \alpha},\tilde\CQ_{I\dot\beta}]				=\delta_{\dot\beta}^{\dot\alpha} \CS_{I}^{\alpha}\cr
[\CP_{\alpha\dot\alpha},\CS_{I}^{\beta}] = -
\delta_{\alpha}^{\beta}\tilde \CQ_{I \dot\alpha}~,\qquad &
[\CP_{\alpha\dot\alpha},\tilde\CS^{I\dot\beta}] =	-\delta_{\dot\alpha}^{\dot\beta} \CQ_{\alpha}^I~.
}}

Using the above commutation relations, one can easily determine the quantum numbers associated with the
various supercharges. These are given in Table~2.

\midinsert
\smallskip
\begintable
            $\CQ$ |$R,r(j_1,j_2)$| $\delta := 2\{\CQ, \CS\}$| Commuting $\delta$s \crthick
 ${\CQ}^1_{+}$  |$\half,\half(\half,0)$| $ E+2j_1-2R-r$
 | $\delta^2_{{-}}$,\quad $\tilde \delta_{{2}\dot{+}}$,\quad $\tilde
           \delta_{{2}\dot{-}}$ \cr
$ {\CQ}^1_{-}$ |$\half,\half(-\half,0)$| $ E-2j_1-2R-r$
 | $\delta^2_{{+}}$,\quad $\tilde \delta_{{2}\dot{+}}$,\quad $\tilde \delta_{{2}\dot{-}}$\cr
$ {\CQ}^2_{ +}$ |$-\half,\half(\half,0)$|
           $ E+2j_1+2R-r$
 | $\delta^1_{{-}}$,\quad $\tilde \delta_{{1}\dot{+}}$,\quad $\tilde \delta_{{1}\dot{-}}$\cr
$  {\CQ}^2_{-}$ | $-\half,\half(-\half,0)$| $ E-2j_1+2R-r$
 | $\delta^1_{{+}}$,\quad $\tilde \delta_{{1}\dot{+}}$,\quad $\tilde \delta_{{1}\dot{-}}$\cr
$\tilde {\CQ}_{{1}\dot{+}}$ |$-\half,-\half(0,\half)$| $ E+2j_2+2R+r$
 | $\tilde \delta_{{2}\dot{-}}$,\quad $\delta^2_{{+}}$,\quad
 $\delta^2_{{-}}$  \cr
   $\tilde {\CQ}_{{1}\dot{-}}$ |$-\half,-\half(0, -\half)$| $ E-2j_2+2R+r$
 | $\tilde \delta_{{2}\dot{+}}$,\quad $\delta^2_{{+}}$,\quad $\delta^2_{{-}}$ \cr
$\tilde {\CQ}_{{2}\dot{+}}$ |$\half,-\half(0,\half)$| $ E+2j_2-2R+r$
 | $\tilde \delta_{{1}\dot{-}}$,\quad
           $\delta^1_{{+}}$,\quad $\delta^1_{{-}}$\cr
$\tilde {\CQ}_{{2}\dot{-}}$ |$\half,-\half(0,-\half)$|$E-2j_2-2R+r$ 
| $\tilde \delta_{{1}\dot{+}}$,\quad $\delta^1_{{+}}$,\quad $\delta^1_{{-}}$ 
\endtable
\Table{2}{List of the various quantum numbers for each supercharge
${\CQ}$.}
\endinsert

The full spectrum of operators can be built by identifying a
``highest-weight state'', which is a superconformal primary of the
$\CN = 2$ SCFT, and acting with the 8 Poincar\'e supercharges to
create superconformal descendants. 
The most generic such (long) multiplet is
denoted by $\CA^E_{R,r(j_1,j_2)}$ and obeys a unitarity
bound. Multiplets that are additionally annihilated by Poincar\'e
supercharges are denoted as short and saturate the bound. The various
shortening conditions can be summarized as
\eqn\shortening{\eqalign{
\CB^I&:\qquad\ [\CQ_\alpha^I,\CO_{\dot\alpha_1\cdots\dot\alpha_{2j_2}}\} =
0~,\qquad {\rm for }~ \alpha = +,-\cr
\bar\CB_I&: \qquad\
[\tilde\CQ_{I\dot\alpha},\CO_{\alpha_1\cdots\alpha_{2j_1}}\} =
0~,\qquad {\rm for }~ \dot\alpha = \dot +,\dot -\cr   
\CC^I&: \cases{\epsilon^{\alpha\beta}[\CQ_\alpha^I,\CO_{\beta\alpha_1\cdots\alpha_{2j_1-1},\dot\alpha_1\cdots\dot\alpha_{2j_2}}\} = 0 ~,& for $j_1> 0$\cr
       \epsilon^{\alpha\beta}\{\CQ_\alpha^I,[\CQ^I_\beta,\CO_{\dot\alpha_1\cdots\dot\alpha_{2j_2}}]\} = 0~,& for
      $j_1= 0$\cr}\cr
\bar\CC_I&:
\cases{\epsilon^{\dot\alpha\dot\beta}[\tilde\CQ_{I\dot\alpha},\CO_{\alpha_1\cdots\alpha_{2j_1},\dot\beta\dot\alpha_1\cdots\dot\alpha_{2j_2-1}}\} = 0~,
& for $j_2 > 0$\cr
       \epsilon^{\dot\alpha\dot\beta}\{\tilde\CQ_{I
      \dot\alpha},[\tilde\CQ_{I\dot\beta},\CO_{\alpha_1\cdots\alpha_{2j_1}}]\} = 0~, & for $j_2= 0$ \cr}\;,
}}
where the operators $\cO$ above are understood to be in the highest
weight representation of $SU(2)_R$.

A full classification of all such possibilities and the associated
short multiplet structure is listed in Table~3.

\midinsert
\smallskip
\tablewidth=13.1cm
\begintable
            Multiplet |\multispan{2}\tstrut\hfil Unitarity
           Bounds\hfil| Shortening \crthick
           $\CA^E_{R,r(j_1,j_2)}$ |$E \ge 2 + 2j_1 +2R + r$
           |$E \ge 2 + 2j_2 +2R - r$ |-- \cr
           $\CB_{R,r(0,j_2)}$| $E = 2R + r$
           |$j_1 = 0 $| $\CB^1$\cr
           $\bar \CB_{R,r(j_1,0)}$| $E= 2R -r$ |$j_2 = 0$ |$\bar \CB_2$\cr
           $\hat \CB_{R}$ |$ E = 2R$| $j_1=j_2=r=0$|$\CB^1\cap\bar\CB_2$\cr
           $\CC_{R,r(j_1,j_2)}$ | $E = 2 + 2j_1 + 2R +r$||$\CC^1$\cr
           $\bar \CC_{R,r(j_1,j_2)}$ | $E = 2 + 2j_2 + 2R -r$||$\bar\CC_2$\cr
           $\CC_{0,r(j_1,j_2)}$ | $E = 2 + 2j_1 +r$|$R = 0$|$\CC^1\cap\CC^2$\cr
           $\bar\CC_{0,r(j_1,j_2)}$ |  $E = 2 + 2j_2-r$|$R = 0 
            $|$\bar\CC_1\cap\bar\CC_2$\cr
           $\hat \CC_{R(j_1,j_2)}$ | $E = 2 + 2R +j_1 + j_2$|$ r= j_2
           - j_1$|$\CC^1\cap\bar\CC_2$\cr
           $\CD_{R(0,j_2)}$ | $E = 1 + 2R +j_2$|$r = j_2 + 1$|$\CB^1\cap\bar\CC_2$\cr
           $\bar \CD_{R(j_1,0)}$ |$E = 1 + 2R +j_1$|$- r = j_1     
           +1$|$\bar\CB_2\cap\CC^1$\cr
           $\CD_{0(0,j_2)}$ |$E = r = 1+ j_2$|$R = 0$|$\CB^1\cap\CB^2\cap\bar\CC_I$\cr
           $\bar\CD_{0(j_1,0)}$ | $E = -r = 1+ j_1$| $R = 
           0$|$\bar\CB_1\cap\bar\CB_2\cap \CC^I$\cr
            $\CE_{r(0,j_2)}$ | $E = r$| $R =0$| $\CB^1\cap\CB^2$\cr 
           $\bar \CE_{r(j_1,0)}$ |$E = -r $| $R = 0$ | $\bar \CB_1\cap\bar\CB_2$ 
\endtable
\Table{3}{The $\CN = 2$ superconformal multiplet structure.}
\endinsert

One conventionally incorporates most of the above into multiplets of type $\CC$, $\bar \CC$, $\hat \CC$ by allowing for
``spin $-{1\over 2}$'' representations as follows \refs{\DolanZH,\BeemSZA}:
\eqn\minushalf{\eqalign{
\CC_{R,r(-\half,j_2)} := \CB_{R + \half, r+\half(0,j_2)},\quad & \bar
\CC_{R,r(j_1, -\half)} := \bar \CB_{R+\half,r - \half(j_1,0)}  \cr
\hat \CC_{R(-\half, j_2 )} := \CD_{R + \half(0,j_2)}, \quad & \hat
\CC_{R(j_1,-\half)} := \bar \CD_{R + \half(j_1,0)}\cr
\hat \CC_{R(-\half, -\half)} := \CD_{R + \half(0,-\half)}& = \bar
\CD_{R+\half(-\half,0)} = \hat \CB_{R + 1}\;.
}}
We make use of this convention in the main part of this paper.

\appendix{B}{RG flow from $(G,G')$ theories}

Here we will explain that all the $(G,G')$ theories defined in \refs{\CecottiFI} are of type C in the sense of subsection~3.6.
Let us first review the definition of $(G,G')$ theories. We consider type IIB string theory on a local Calabi--Yau threefold given by the hypersurface singularity
\eqn\singularity{
W(x_1,\cdots,x_4) = 0~,
}
in ${\bf C}^4$. Here $W(x_1,\cdots,x_4)$ is a quasi-homogeneous polynomial such that
\eqn\scaling{
W(\zeta^{q_i} x_i) = \zeta W(x_i)
}
for all $\zeta\in {\bf C}^*$.
We also assume that $dW=0$ if and only if $x_i=0$. The holomorphic 3-form on the threefold is given by
$\Omega = {dx_1  dx_2 dx_3 dx_4 \over dW}$ up to rescaling.
Since the Calabi--Yau three-fold is non-compact, it engineers an $\CN = 2$ gauge theory in the transverse four dimensions. If the singularity \singularity\ arises at finite distance in the moduli space of some compact Calabi--Yau threefold, we can think of the 4d gauge theory as a rigid supersymmetry limit of some 4d supergravity. In terms of $\hat{c} \equiv \sum_{i=1}^4(1-2q_i)$, the necessary and sufficient condition for this is written as $2-\hat{c} > 0$ \GukovYA. A class of $W(x_i)$ satisfying this condition is given by \CecottiFI
\eqn\GG{
W(x_1,\cdots,x_4) = W_G(x_1,x_2) + W_{G'}(x_3,x_4)~,
}
where $G,G'=A_n,D_n,E_n$ and
\eqn\hoge{\eqalign{
W_{A_n}(x,y) &= x^{n+1} + y^2~,\cr
W_{D_n}(x,y) &= x^{n-1} + xy^2~,\cr
W_{E_6}(x,y) &= x^3 + y^4~,\cr
W_{E_7}(x,y) &= x^3 + xy^3~,\cr
W_{E_8}(x,y) &= x^3 + y^5~.\cr
}}
The 4d theory engineered by the singularity \GG\ is called the $(G,G')$ theory \CecottiFI.

The BPS states of the 4d theory correspond to D3-branes wrapping on supersymmetric compact 3-cycles of the Calabi--Yau threefold. Since all the compact 3-cycles are now shrinking at $x_i=0$, the 4d BPS states are all massless. This suggests that the $(G,G')$ theory is a 4d superconformal filed theory of Argyres--Douglas type \refs{\ArgyresXN, \ArgyresJJ}.
The transformation $x_i\to \zeta^{q_i}x_i$, which keeps the equation $W(x_i)=0$ invariant, corresponds to a scale transformation in four dimensions. To identify the scaling dimension of the variable $x_i$, let us consider a 4d particle corresponding to a D3-brane wrapping on a 3-cycle of the Calabi--Yau threefold. The central charge of the 4d particle is given by $\oint \Omega$, where the integration is taken over the 3-cycle.  Now $x_i\to \zeta^{q_i}x_i$ induces
\eqn\hogehoge{
\oint \Omega \;\to\; \zeta^{{2-\hat{c} \over 2}} \oint  \Omega~.
}
Since the central charge has scaling dimension one, we interpret $\xi :=\zeta^{{2-\hat{c} \over 2}}$ as the dilatation factor of a 4d scale transformation. Then $\zeta^{q_i} = \xi^{2q_i/(2-\hat{c})}$ implies that  the scaling dimension of $x_i$ is written as \ShapereXR
\eqn\mumumu{
[x_i] = {2q_i\over 2-\hat{c}}~.
}

\subsec{Deformation of the singularity}

The deformations of the singularity \singularity\ correspond to the complex structure moduli of the Calabi--Yau threefold. In particular, complex structure moduli associated with the vanishing 3-cycles govern the Coulomb branch of the 4d gauge theory.
Varying such moduli replaces $W(x_i)=0$ with
\eqn\deformation{
W(x_i) +  \delta W(x_i)=0~,
}
for some lower order polynomial $\delta W(x_i)$. To find a basis of such deformations, let us consider the ring
\eqn\ring{
\CR : = {\bf C}[x_1,\cdots,x_4] / I_W~,
}
where $I_W$ is an ideal of ${\bf C}[x_1,\cdots,x_4]$ generated by all $\partial W/\partial x_i$. It is known \refs{\Arnold, \ShapereXR} that the number of independent vanishing 3-cycles is equal to ${\rm dim}_{\bf C}\,\CR$ (as a vector space over ${\bf C}$). Let us take a set of polynomials $\{W_k\}$ which represent a basis of $\CR$. Then the most general deformation is written as
\eqn\basis{
W(x_i) + \sum_{k=1}^{{\rm dim}_{\bf C} \CR}c_k W_k(x_i) = 0
}
for $c_k\in{\bf C}$. We can think of $\{c_k\}$ as a local coordinate
of the complex structure moduli near the singular point. In terms of
4d physics, they are interpreted as masses and Coulomb branch
parameters of the engineered gauge theory. In particular, $c_k=0$ corresponds to the superconformal point.

Let us take $W_k(x_i)$ to be monomials in $x_i$. There exists $Q_k\in{\bf Q}^+$ such that 
\eqn\charge{
W_k(\zeta^{q_i}x_i) = \zeta^{Q_k}W_k(x_i)
}
for all $\zeta \in {\bf C}^*$. The scaling dimension of $c_k$ in 4d
physics is then evaluated to be \ShapereXR
\eqn\dimensions{
[c_k] = {2(1-Q_k)\over 2-\hat{c}}.
}
For any $c_k$ there exists $c_\ell$ such that $[c_k] + [c_\ell] = 2$, which is consistent with the arguments of \ArgyresXN. If $[c_k]<1$, the parameter $c_k$ is interpreted as a coupling constant of the theory. The corresponding deformation of the theory is given by
\eqn\coupling{
\int d^4\theta\; c_k\CO_k~,
}
with some Coulomb branch operator $\CO_k$ of dimension $2-[c_k] = [c_\ell]$. Now $c_\ell$ is interpreted as the vev of the operator $\CO_k$. On the other hand, if $[c_k]>1$, the roles of $c_k$ and $c_\ell$ are exchanged.
If $[c_k]=1$, the parameter $c_k$ is interpreted as a mass deformation parameter.

\subsec{Reduction to Seiberg--Witten description}

If the polynomial $W(x_i)$ is of the form
\eqn\SW{
W(x_i) = x_1^2 + x_2^2 + \tilde{W}(x_3,x_4)~,
}
the Coulomb branch of the 4d theory has a Seiberg--Witten description
with the curve $\tilde{W}(s,t) = 0$ and the 1-form $\lambda = sdt$ on
it \CecottiFI. The reason for this is that any monomial containing
$x_1$ or $x_2$ becomes trivial in $\CR$; there is no non-trivial
complex structure deformation in the $x_{1,2}$ directions. For
example, any $(A,A)$ theory has a Seiberg--Witten description. These
types of theories were studied in \XieHS\ from the viewpoint of
M5-branes on a Riemann surface, while their RG-flows investigated in
\XieJC. On the other hand, if $W(x_i)$ cannot be expressed in the form
\SW\ (even after changing variables), the corresponding 4d theory does
not seem to have a Seiberg--Witten description \CecottiFI. For
instance, the $(D_n,D_m)$ theories for $n,m\geq 4$ and $(E,E)$ theories
provide such examples.

\subsec{RG flow by relevant deformations}

We now study RG flows from $(G,G')$ theories to show that they are of type C in the sense of subsection~3.6. Let us start with the $(A_n,A_m)$ theory. The corresponding Calabi--Yau singularity is described by
\eqn\ANcurve{
W(x,y,s,t)= x^{n+1} + y^2 + s^{m+1} + t^2 = 0~.
}
We have $q_x = {1\over n+1}, q_s={1\over m+1}, q_y=q_t={1\over 2}$, and therefore $2-\hat{c}=2({1\over n+1}+{1\over m+1})$. The dimensions of the variables are
\eqn\dimx{
[x]={1\over 1+{n+1\over m+1}}~,\qquad [s]={1\over 1+{m+1\over n+1}}~,\qquad [y] = [t]={1 \over 2({1\over n+1}+{1\over m+1})}~.
}
The ring $\CR = {\bf C}[x,y,s,t]/I_W$ is now
${\bf C}[x,y,s,t]/\langle x^n,y,s^m,t \rangle$.
Since there is no complex structure deformation involving $y$ or $t$, we have a Seiberg--Witten description with the curve
\eqn\SWAn{
x^{n+1} + s^{m+1} = 0~.
}
and the 1-form $\lambda = sdx$.
The most general deformed curve in the Coulomb branch is written as
\eqn\deformedAn{
x^{n+1} + s^{m+1} + \sum_{k=0}^{n-1}\sum_{\ell=0}^{m-1} c_{k,\ell}x^{k}s^{\ell}=0~.
}
The dimensions of the parameters are
\eqn\dimcoef{
[c_{k,\ell}] = {(m+1)(n+1) - k(m+1)-\ell(n+1) \over m+n+2}~,
}
which satisfy $[c_{k,\ell}]+[c_{n-k-1,m-\ell-1}]=2$~.

In particular,
\eqn\dimpair{
[c_{n-1,0}] = {2(m+1) \over m+n+2}~,\qquad [c_{0,m-1}] = {2(n+1)\over m+n+2}~.
}
If $n\geq m$, $0<[c_{n-1,0}] \leq 1$ implies that $c_{n-1,0}$ is a relevant coupling or mass parameter. The corresponding deformation does not turn on any vev of the Coulomb branch operators, and leads to the deformed curve
\eqn\deformation{
x^{n+1} + s^{m+1} + c_{n-1,0}\,x^{n-1} = 0~.
} 
We now have several massive BPS states whose masses are proportional to $(c_{n-1,0})^{1/ [c_{n-1,0}]}$. In the deep IR, all the massive degrees of freedom are integrated out. This corresponds to focusing on the region $|x|\ll \sqrt{|c_{n-1,0}|}$. Then the IR curve is written as
\eqn\deformedAn{
x^{n-1} + s^{m+1} = 0~,
}
after rescaling $x$ and $s$.\foot{To be more precise, we rescale $x$ and $s$ so that $c_{n-1,0}x^{n-1}+s^{m+1} = 0$ becomes $x^{n-1}+s^{m+1}=0$ while keeping $\lambda  =sdx$ fixed. This changes the scaling dimensions of $x$ and $s$.}
Since this is the curve of the $(A_{n-2}, A_m)$ theory, we identify that the IR CFT is the $(A_{n-2},A_m)$ theory (there is also a decoupled $U(1)$). Thus, in the case $n\geq m$, the $(A_n,A_m)$ theory can flow to the $(A_{n-2},A_m)$ theory via a relevant perturbation. 
On the other hand, if $n\leq m$, $c_{0,m-1}$ is a relevant coupling or mass parameter. The corresponding relevant deformation leads to the $(A_n,A_{m-2})$ theory (and a decoupled $U(1)$) in the infrared.

By repeating this procedure, the $(A_n,A_m)$ theory eventually flows (up to several decoupled $U(1)$ factors) to the $(A_1,A_1)$, $(A_1,A_0)$ or $(A_2,A_0)$ theory via relevant deformations. Here the $(A_1,A_0)$ and $(A_2,A_0)$ theories are trivial theories while the $(A_1,A_1)$ theory is a theory of a single free hypermultiplet. Therefore any $(A_n,A_m)$ theory can flow to a Lagrangian theory via relevant deformations. 
This means that all the $(A_n,A_m)$ theories are of type C in the sense of subsection~3.6.

It is straightforward to generalize this argument to the other $(G,G')$ theories. For example, the $(A_n,E_6)$ theory can flow to some $(A_m,D_4)$ theory without turing on vev's of the Coulomb branch operators. Then the latter can further flow to some $(A,A)$ theory via relevant deformations. Since all the $(A,A)$ theories are of type C, the original $(A_n,E_6)$ theory is also of type C. Exactly the same argument shows that all the $(G,G')$ theories are of type C.
Our discussion in subsection~3.6 then implies that  $(G,G')$ theories have no exotic chiral operators.

\listrefs

\end